\documentclass[twocolumn,twoside]{IEEEtran}
\usepackage{amsmath,amsfonts}
\usepackage{array}
\usepackage[caption=false,font=normalsize,labelfont=sf,textfont=sf]{subfig}
\usepackage{textcomp}
\usepackage{stfloats}
\usepackage{url}
\usepackage{verbatim}
\usepackage{graphicx}
\usepackage[noadjust]{cite}
\usepackage{graphicx}
\usepackage{amsthm}
\usepackage[table]{xcolor}
\usepackage{algorithm}
\usepackage{algpseudocode}
\usepackage{tikz}
\usepackage{amssymb}
\usepackage{mathtools}
\usepackage{overpic}

\DeclarePairedDelimiter\abs{\lvert}{\rvert}

\DeclarePairedDelimiter\ceil{\lceil}{\rceil}
\DeclarePairedDelimiter\floor{\lfloor}{\rfloor}
\DeclarePairedDelimiter\parenv{\lparen}{\rparen}
\DeclarePairedDelimiter\sparenv{\lbrack}{\rbrack}

\DeclarePairedDelimiter\set{\{}{\}}

\theoremstyle{plain}
\newtheorem{theorem}{Theorem}
\newtheorem{lemma}[theorem]{Lemma}

\newtheorem{definition}[theorem]{Definition}
\newtheorem{example}[theorem]{Example}

\newtheorem{construction}{Construction}

\renewcommand{\leq}{\leqslant}

\renewcommand{\geq}{\geqslant}

\newcommand{\bfc}{\mathbf{c}}
\newcommand{\bfd}{\mathbf{d}}
\newcommand{\bfe}{\mathbf{e}}

\newcommand{\bfx}{\mathbf{x}}
\newcommand{\bfy}{\mathbf{y}}

\newcommand{\bzero}{\mathbf{0}}

\newcommand{\Z}{\mathbb{Z}}

\newcommand{\eqdef}{\triangleq}

\begin{document}

\title{Improved Constructions of Skew-Tolerant Gray Codes}

\author{
Gabriel Sac Himelfarb and Moshe~Schwartz,~\IEEEmembership{Fellow,~IEEE}%
\thanks{This paper will be submitted in part to the IEEE International Symposium on Information Theory 2025.}%
\thanks{Gabriel Sac Himelfarb is with the Department of Electrical and Computer Engineering, McMaster University, Hamilton, ON, L8S 4K1, Canada (e-mail: sachimeg@mcmaster.ca).}%
\thanks{Moshe Schwartz is with the Department of Electrical and Computer Engineering, McMaster University, Hamilton, ON, L8S 4K1, Canada, and on a leave of absence from the School
   of Electrical and Computer Engineering, Ben-Gurion University of the Negev,
   Beer Sheva 8410501, Israel
   (e-mail: schwartz.moshe@mcmaster.ca).}
}

\maketitle

\begin{abstract}
We study skew-tolerant Gray codes, which are Gray codes in which changes in consecutive codewords occur in adjacent positions. We present the first construction of asymptotically non-vanishing skew-tolerant Gray codes, offering an exponential improvement over the known construction. We also provide linear-time encoding and decoding algorithms for our codes. Finally, we extend the definition to non-binary alphabets, and provide constructions of complete $m$-ary skew-tolerant Gray codes for every base $m\geq 3$.
\end{abstract}


\section{Introduction}
\IEEEPARstart{E}{ver} since their introduction~\cite{Gra53}, Gray codes have been of great interest both for their practical applications in data communications and storage, and for the theoretical problems that arise from their study. In their original forms, Gray codes are listings of all binary $n$-tuples, without repetitions, and with the added restrictions that consecutive tuples differ by a single bit flip. Multiple variants have been extensively investigated, and are still today, such as snake-in-the-box codes~\cite{Kau58,Sin66,AbbKat91}, circuit codes~\cite{Kle67,PatTul98,HilPat01,HooRecSawWon15}, and single-track Gray codes~\cite{HilPatBra96,EtzPat96b,SchEtz99,HilPat01,HooRecSawWon15}, to name a few. Moreover, Gray codes have inspired the idea of finding listings of combinatorial objects such that there is only a small change from one element to the next. They have thus been extended to codes over subspaces~\cite{Sch14a,Bra14}, permutations~\cite{YehSch12b,HorEtz14,ZhaGe16a,Hol17,ZhaGe16b,WanFu20,YehSch17}, and a multitude of other structures. We refer the reader to the classical survey~\cite{Sav97} and the more recent survey~\cite{Mut22} for a comprehensive review of Gray codes. 

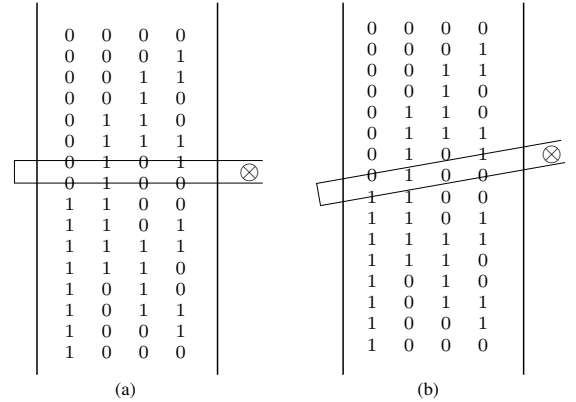
\begin{figure}[t]
\begin{center}
    \begin{overpic}[scale=0.3]
      {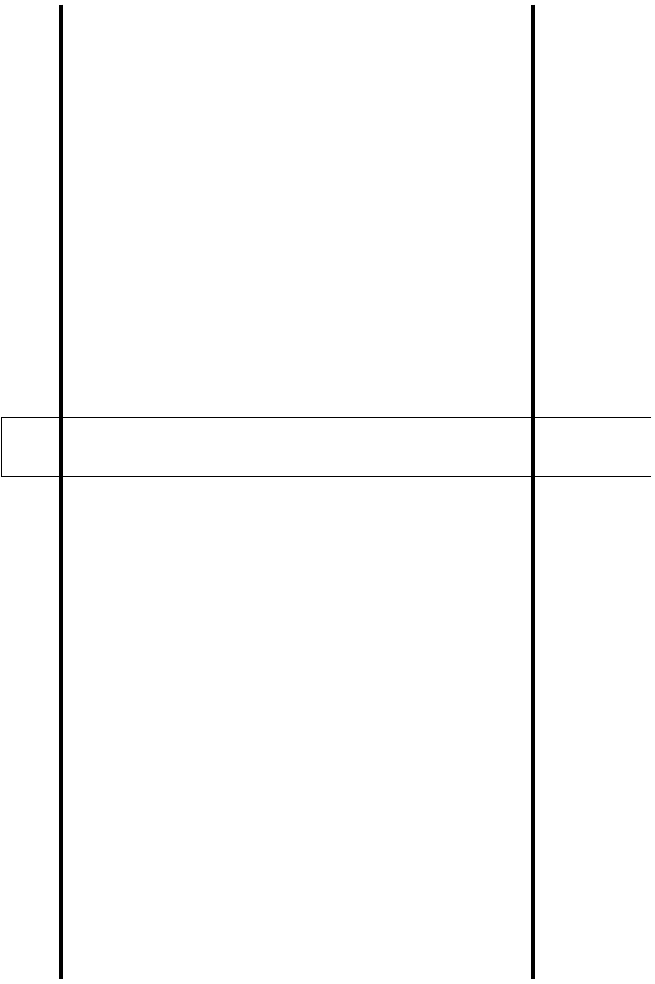}
      \put(60,52.5){$\otimes$}
      \put(27,-5){\scriptsize{(a)}}
      \put(10,47){\scriptsize $\begin{array}{cccc}
          0 & 0 & 0 & 0 \\
          0 & 0 & 0 & 1 \\
          0 & 0 & 1 & 1 \\
          0 & 0 & 1 & 0 \\
          0 & 1 & 1 & 0 \\
          0 & 1 & 1 & 1 \\
          0 & 1 & 0 & 1 \\
          0 & 1 & 0 & 0 \\
          1 & 1 & 0 & 0 \\
          1 & 1 & 0 & 1 \\
          1 & 1 & 1 & 1 \\
          1 & 1 & 1 & 0 \\
          1 & 0 & 1 & 0 \\
          1 & 0 & 1 & 1 \\
          1 & 0 & 0 & 1 \\
          1 & 0 & 0 & 0 \\
          \end{array}$}
    \end{overpic}
    \hspace{3ex}
    \begin{overpic}[scale=0.3]
      {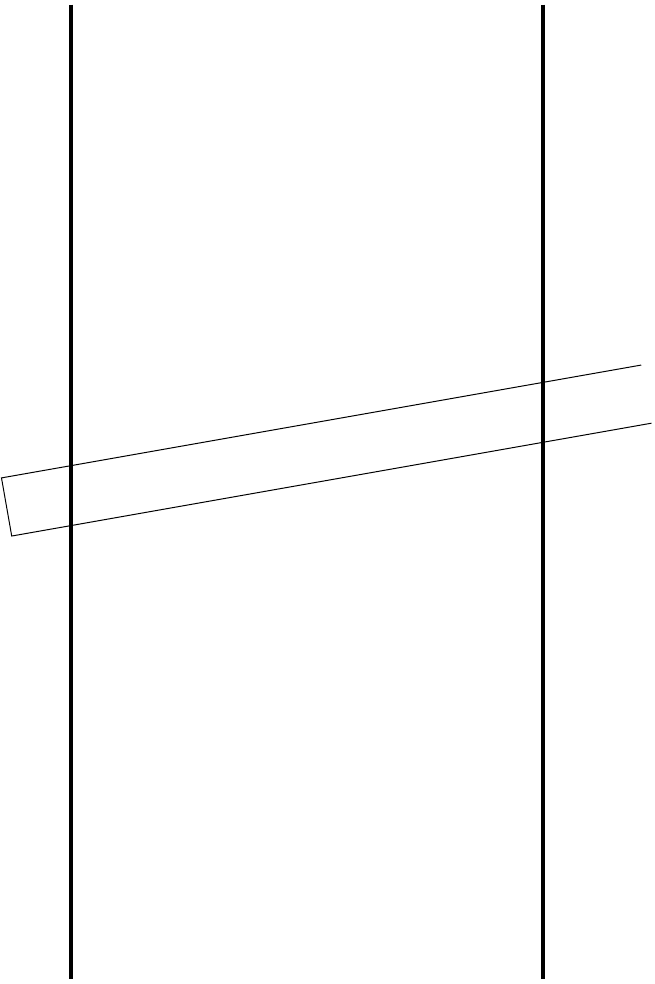}
      \put(60,57.5){$\otimes$}
      \put(27,-5){\scriptsize{(b)}}
      \put(10,49){\scriptsize $\begin{array}{cccc}
          0 & 0 & 0 & 0 \\
          0 & 0 & 0 & 1 \\
          0 & 0 & 1 & 1 \\
          0 & 0 & 1 & 0 \\
          0 & 1 & 1 & 0 \\
          0 & 1 & 1 & 1 \\
          0 & 1 & 0 & 1 \\
          0 & 1 & 0 & 0 \\
          1 & 1 & 0 & 0 \\
          1 & 1 & 0 & 1 \\
          1 & 1 & 1 & 1 \\
          1 & 1 & 1 & 0 \\
          1 & 0 & 1 & 0 \\
          1 & 0 & 1 & 1 \\
          1 & 0 & 0 & 1 \\
          1 & 0 & 0 & 0 \\
          \end{array}$}
    \end{overpic}
\end{center}
\caption{Reading heads mounted over a surface (a) without a skew, and (b) with a skew.}
\label{fig:disk}
\end{figure}

In~\cite{WilBla06}, Wilson and Blaum defined skew-tolerant Gray codes as Gray codes such that changes in consecutive codewords occur in adjacent positions. Their motivation behind this definition was to allow for better performance in mechanical encoder systems where the reading head might be skewed. In Figure~\ref{fig:disk}(a), a surface contains distinct binary vectors of length $4$ written as rows of a matrix. Four sensors are mounted on a reading head (denoted by $\otimes$), and the head moves vertically. By reading the binary vector beneath it, it can deduce its position. However, if it attempts to do so in-between two rows, each sensor may independently read the bit below or above it. If more than one bit is in question, the resulting reading may be neither the vector immediately below the head's position, nor the one immediately above it. This may result in a location error. To avoid this, the rows form a binary Gray code, namely, a single bit changes between any two consecutive rows. Thus, even if getting a reading in-between rows, the resulting reading is the entire vector immediately above or below the reading-head location. When the reading head is mounted with a skew, even if we use a (general) binary Gray code, more than one sensor may see a bit change at some moment, as seen in Figure~\ref{fig:disk}(b). As~\cite{WilBla06} shows, by using a skew-tolerant Gray code, the skew angle at which the system operates without any location error is maximized.

Another major advantage of skew-tolerant Gray codes is that they can be easily compressed, as they are completely determined by the first codeword, the position of the first bit flip, and a sequence of directions (right, left) that indicate the next change position. Thus, only $\Theta(1)$ bits per codeword are needed for storage.

Wilson and Blaum's definition is equivalent to the $\overline{P_n}$-compatible Gray codes defined previously by Slater in~\cite{Sla79} and~\cite{Sla89}, where $P_n$ represents the path graph of $n$ vertices. Several other works have studied Gray codes compatible with more general graphs, like trees with infinite diameter \cite{WilErn02}, hypercubes ~\cite{DimDvoTomGreSkr09} and complete multipartite graphs ~\cite{BulRus96}.

The only attempt at constructing large skew-tolerant Gray codes that we know of is that of~\cite{WilBla06}. The size of the known codes of $n$ bits is of the order of $(\sqrt{3})^n$ codewords, which is exponentially smaller than the theoretical bound of $2^n$ codewords, namely, a complete code. Slater~\cite{Sla89} conjectured that no complete skew-tolerant ($\overline{P_n}$-compatible) Gray code exists for $n\geq 7$. This leaves open the question: what is the largest possible skew-tolerant Gray code with length of $n$ bits?

The main contribution of this work is to present a construction of asymptotically non-vanishing skew-tolerant Gray codes. More precisely, we present codes of length $n$ and size $\approx c\cdot 2^n$ with $c$ a constant that depends on the parity of $n$. For $n$ even, $c>0.76$, and for $n$ odd $c>0.84$. Thus, our codes offer an exponential improvement over the codes of~\cite{WilBla06}. We also study the generalization of these codes to larger alphabets, and show that for any alphabet size $m\geq 3$, there are complete skew-tolerant Gray codes.

This work is organized as follows: in Section~\ref{sec:preliminaries} we introduce the definition of $k$-skew-tolerant Gray codes ($k$-SkTGCs), as well as necessary notations used throughout the paper. In Section~\ref{sec:binary} we start by studying $3$-SkTGCs and $2$-SkTGCs, showing that complete codes are possible. By using a similar approach, we then construct asymptotically non-vanishing $1$-SkTGCs. We also provide linear-time encoding and decoding algorithms for all constructed binary codes. Finally, in Section~\ref{sec:nonbinary} we generalize the definition of skew-tolerant Gray codes to non-binary alphabets, and we show that it is always possible to construct complete $m$-ary skew-tolerant Gray codes of any length. 

\section{Preliminaries}
\label{sec:preliminaries}

We begin by introducing necessary notation and definitions. Throughout the paper we assume a finite alphabet, which w.l.o.g., may be taken to be $\Z_m$, the ring of integers modulo $m\geq 2$. We use lower-case letters to denote scalars, bold lower-case letters to denote vectors, and upper-case letters to denote sequences of vectors.

Let $\bfx=(x_1,x_2,\dots,x_n)\in\Z_m^n$ be a vector. Most of the time we index the entries of vectors starting from $1$. When we deviate from this norm we shall write so explicitly. We denote by $\bfx[i]\eqdef x_i$ the entry at position $i$ of $\bfx$, and $\bfx[i\dots  j]\eqdef (x_i,x_{i+1},\dots,x_j)$ the entries of $\bfx$ from position $i$ to $j$. The $i$-th standard unit vector, denoted $\bfe_i$, is the vector all of whose entries are $0$ except for position $i$ which contains $1$. We also use $\bzero$ to denote the all-zero vector. When writing specific vectors, we shall often conveniently omit parentheses and commas.

Consider now a sequence of vectors, $C=\bfc_0,\bfc_1,\dots,\bfc_{P-1}$, where for all $i$, $\bfc_i\in\Z_m^n$. We say this sequence has length $n$ and size $\abs{C}=P$. For conveniently handling cyclic sequences, we index the vectors in the sequence starting from $0$. We shall also find it easier to depict such a sequence by a $P\times n$ array whose $i$-th row is $\bfc_i$, namely,
\[
C=\begin{bmatrix} \bfc_0 \\ \bfc_1 \\ \vdots \\ \bfc_{P-1} \end{bmatrix}.
\]
Simple operations on $C$, which remove either the first or last vector (or both), or reverse their order, are defined as follows:
\begin{align*}
C'&\eqdef\begin{bmatrix} \bfc_1 \\ \bfc_2 \\ \vdots \\ \bfc_{P-1} \end{bmatrix}, & 
C^*&\eqdef\begin{bmatrix} \bfc_0 \\ \bfc_1 \\ \vdots \\ \bfc_{P-2} \end{bmatrix}, \\
\widehat{C}&\eqdef\begin{bmatrix} \bfc_1 \\ \bfc_2 \\ \vdots \\ \bfc_{P-2} \end{bmatrix}, & 
\overleftarrow{C}&\eqdef\begin{bmatrix} \bfc_{P-1} \\ \bfc_{P-2} \\ \vdots \\ \bfc_0 \end{bmatrix}.
\end{align*}

We can now define Gray codes.

\begin{definition}
\label{def:gray}
An \emph{$m$-ary Gray code} of length $n$ and size $P$ is a sequence of $P$ distinct vectors (codewords) of length $n$ over $\Z_m$, $C=\bfc_0,\bfc_1,\dots,\bfc_{P-1}$, where $\bfc_i\in\Z_m^n$, such that for all $0\leq i\leq P-2$, $\bfc_i$ and $\bfc_{i+1}$ differ by a single bit flip, i.e.,
\[
\bfc_{i+1}-\bfc_i = \pm\bfe_{\delta_i},
\]
for some $1\leq \delta_i\leq m$. If the latter property also holds for $\bfc_{P-1}$ and $\bfc_0$, namely,
\[
\bfc_0-\bfc_{P-1} = \pm\bfe_{\delta_{P-1}},
\]
we say the Gray code is \emph{cyclic}. If $P=m^n$, we say the Gray code is \emph{complete}.
\end{definition}

We note that some papers relax the above definition, allowing an arbitrary change in a single position when moving from a codeword to the next, whereas we use the stricter definition whereby that change is a $\pm 1$ in the value of the position. This distinction is moot for binary and ternary Gray codes.

\begin{definition}
\label{def:trans}
Let $C=\bfc_0,\bfc_1,\dots,\bfc_{P-1}$ be an $m$-ary Gray code of length $n$ and size $P$. Let $\delta_i$ be as given in Definition~\ref{def:gray}. The \emph{transition sequence} of $C$ is defined as $\delta_0,\delta_1,\dots,\delta_{P-2}$. If $C$ is cyclic, we define its transition sequence to also include the change location from $\bfc_{P-1}$ to $\bfc_0$, namely, it is $\delta_0,\delta_1,\dots,\delta_{P-1}$.
\end{definition}

\begin{example}
Consider the following binary Gray code of length $3$ and size $8$:
\[
C=\begin{bmatrix}
000\\
100\\
110\\
010\\
011\\
111\\
101\\
001
\end{bmatrix}.
\]
It is a complete cyclic Gray code, and its transition sequence is $1,2,1,3,1,2,1,3$.
\end{example}

Definition~\ref{def:gray} is the starting point for many variants of Gray codes known in the literature. Depending on the extra requirements, a complete Gray code is not always possible. Assume we have a parametric construction for $m$-ary Gray codes resulting in a sequence of codes, $C_1, C_2,\dots$, where $C_i$ has length $n_i$ and size $P_i$. Further assume that $\lim_{i\to\infty}n_i=\infty$. To measure the efficiency of the construction we say it is \emph{asymptotically non-vanishing} if
\[
\lim_{i\to\infty} \frac{P_i}{m^{n_i}}=c,
\]
for some real $0<c\leq 1$. An exponentially weaker requirement is \emph{asymptotically constant rate}, defined by
\[
\lim_{i\to\infty} \frac{\log_m(P_i)}{n_i}=\rho,
\]
for some real $0<\rho\leq 1$.

We can now define the object we shall study throughout the paper, skew-tolerant Gray codes.

\begin{definition}\label{def:kSkTGC}
An $m$-ary $k$-skew-tolerant Gray code ($k$-SkTGC\footnote{While~\cite{WilBla06} used the acronym STGC to refer to a skew-tolerant Gray code, we prefer to use SkTGC to avoid confusion with the acronym for single-track Gray codes.}) of length $n$ and size $P$ is a cyclic Gray code (of the same parameters) with transition sequence $\delta_0,\delta_1,\dots, \delta_{P-1}$ such that 
\[\abs*{\delta_i-\delta_{i+1}}\leq k,\]
where indices are considered modulo $M$.
\end{definition}

We note that $k$-SkTGCs are cyclic by definition. However, we shall occasionally need non-cyclic $k$-SkTGCs as component codes for the cyclic code construction. In that case we shall explicitly write \emph{non-cyclic} $k$-SkTGCs.

Definition~\ref{def:kSkTGC} extends the definition given in~\cite{WilBla06} both by allowing $m$-ary alphabets, as well as by relaxing the skew-tolerance condition to allow for consecutive changes which are $k$-close but not necessarily at distance $1$. Thus, a binary $1$-SkTGC of Definition~\ref{def:kSkTGC} is exactly a skew-tolerant Gray code as defined in~\cite{WilBla06}. The larger the value of $k$, the more prone to skew-related reading errors a device will be (the maximum angle at which a sensor may be allowed to be will decrease, see~\cite{WilBla06}).

We emphasize that the generalization from SkTGCs to $k$-SkTGCs is not arbitrary. Apart from a pure mathematical interest, the constructions of complete binary $3$-SkTGCs and $2$-SkTGCs we provide in Section~\ref{sec:binary}, pave the way to our construction of  asymptotically non-vanishing binary $1$-SkTGCs, thereby exponentially improving upon the best construction of~\cite{WilBla06} that only has an asymptotically constant rate of $\frac{\log_2(3)}{2}\approx 0.792$. Additionally, from an application point of view, the complete $3$-SkTGCs and $2$-SkTGCs trade off skew tolerance for a complete Gray code. Finally, the complete binary $2$-SkTGC we construct is used to construct a complete quaternary $1$-SkTGC in Section~\ref{sec:nonbinary}.

Another interesting property of a Gray code is its induced graph, which we now define. The \emph{induced graph} of a cyclic Gray code of length $n$, size $P$, and transition sequence $\delta_0,\dots  ,\delta_{P-1}$, is the undirected graph with $n$ vertices $1, 2, \dots  , n$, and edges $\set{\delta_i,\delta_{i+1}}$ for $0\leq i\leq P-1$, where indices are taken modulo $P$. Following~\cite{WilErn02}, given an undirected graph $G$ with vertices numbered $1, \dots , n$, we say a cyclic Gray code of length $n$ is $\overline{G}$-\emph{compatible} if its induced graph is a spanning subgraph of $G$. Similarly, the term $G$-compatible is reserved for Gray codes when viewed as non-cyclic.

We can now reformulate Definition~\ref{def:kSkTGC}. Let $P^\circ_n$ denote the path graph of $n$ vertices with self loops:
\begin{center}
    \begin{tikzpicture}[every loop/.style={}]
    \node (1) at (0,0) {1};
    \node (2) at (1,0) {2};
    \node (x) at (2,0) {};
    \node (n) at (3.5,0) {$n$};
    \node (n-1) at (2.5,0) {$n-1$};
    \draw (1) -- (2);
    \path
    (1) edge [loop above] node {} (1);
    \path
    (2) edge [loop above] node {} (2);
    \path
    (n) edge [loop above] node {} (n);
        \path
    (n-1) edge [loop above] node {} (n-1);
    \draw (n-1) -- (n);
    \path (2) -- node[auto=false]{\ldots} (x);
\end{tikzpicture}
\end{center}
Then, an $m$-ary $1$-SkTGC of length $n$ is a $\overline{P^\circ_n}$-compatible Gray code. If we restrict ourselves to the binary case, and assume the code has at least three codewords, then two consecutive changes cannot be in the same position (otherwise a codeword would repeat itself). Thus, a binary $1$-SkTGC of length $n$ with at least three codewords is a $\overline{P_n}$-compatible Gray code, where $P_n$ is the path graph on $n$ vertices:
\begin{center}
    \begin{tikzpicture}[every loop/.style={}]
    \node (1) at (0,0) {1};
    \node (2) at (1,0) {2};
    \node (x) at (2,0) {};
    \node (n) at (3.5,0) {$n$};
    \node (n-1) at (2.5,0) {$n-1$};
    \draw (1) -- (2);
    \draw (n-1) -- (n);
    \path (2) -- node[auto=false]{\ldots} (x);
\end{tikzpicture}
\end{center}
More generally, an $m$-ary $k$-SkTGC of length $n$ is a $\overline{T_{k,n}}$-compatible Gray code, where $T_{k,n}$ is Toeplitz graph\footnote{Recall that a Toeplitz graph is a graph whose adjacency matrix is a symmetric matrix which is constant along its diagonals.} of $n$ vertices with adjacency matrix whose first row is 
\[\sum_{i=1}^{k+1}\bfe_i=(\underbrace{1,\dots , 1}_{k+1}, 0,\dots  ,0).\]

To the best of our knowledge, the literature on graphs induced by Gray codes has only considered complete binary codes. Slater~\cite{Sla79,Sla89} was the first to consider the question of $P_n$-compatible complete binary Gray codes. He also raised the general question: which graphs can be induced by a complete binary Gray code? It is known (by computer search) that there are $\overline{P_n}$-compatible complete binary Gray codes for $n=2,3,5$, but such codes do not exist for $n=4,6,7$. If we consider the codes to be non-cyclic, then there are such codes for $n=4,6$. Slater conjectured in~\cite{Sla89} that no $P_n$-compatible complete binary Gray code exists for $n\geq 7$, but the problem is still open. The only attempt that we know of to build $\overline{P_n}$-compatible and $P_n$-compatible binary Gray codes (even if incomplete) is the recursive construction of Wilson and Blaum in~\cite{WilBla06}, which attains asymptotic rate $\frac{\log_2(3)}{2}\approx 0.792$.

For completeness, we mention some other results on the question of which graphs are induced by complete binary Gray codes. Wilmer and Ernst~\cite{WilErn02} constructed codes whose induced graph is a tree of arbitrarily large diameter, and proved that there is a certain family of trees such that no code is compatible with them. Other results include the existence of codes which are compatible with complete multipartite graphs~\cite{BulRus96}, and codes which induce complete graphs~\cite{SupVan08}, complete bipartite graphs~\cite{Sup17}, and the hypercube~\cite{DimDvoTomGreSkr09}. This last one allows for compressed storage of $\Theta(\log \log n)$ bits per $n$-bit codeword, to be compared with the $\Theta(1)$ bits per $n$-bit codeword for $k$-SkTGC (assuming $k$ is constant).

\section{Binary skew-tolerant Gray codes}\label{sec:binary}

In this section we present constructions for binary SkTGCs. Our main objective is to give a construction of large $1$-SkTGCs. To this end, we first show that a particular case of a known Gray code construction, the so-called \emph{supercomposite Gray codes} (see~\cite{WilErn02}), is in fact a complete $3$-SkTGC. We then improve on this construction to obtain complete $2$-SkTGCs. Finally, we obtain asymptotically non-vanishing $1$-SkTGCs with a similar but more complex construction.

\subsection{The case $k=3$}

The family of supercomposite Gray codes was introduced in~\cite{WilErn02} in order to disprove a conjecture by Bultena and Ruskey. These codes generalize the binary reflected Gray codes, whose recursive construction uses a reflection, by introducing a shift operation as well. We show that a specific choice of parameters results in complete $3$-SkTGCs.

\begin{construction}\label{const:skip3}
    Consider the code of length $1$ 
    \[
    A_1 \eqdef \begin{bmatrix} 0\\1 \end{bmatrix},
    \]
    and for every $n\geq 1$, define the code $A_{n+1}$ as
    \[
    A_{n+1} \eqdef
    \sparenv*{
  \begin{array}{c|c}
    0\dots 0 & 0 \\
    \hline
     & 1 \\
    A_n & \vdots \\
     & 1 \\
     \hline
     & 0 \\
    \overleftarrow{A_n'} & \vdots \\
     & 0 
    \end{array}
    }
    \begin{array}{l}
    \text{ending of 2nd block} \\
    \phantom{1}\\
    \text{1st block}\phantom{\vdots}\\
    \phantom{1}\\
    \phantom{1}\\
    \text{2nd block}\phantom{\vdots}\\
    \phantom{1}\\
    \end{array}
    \]
\end{construction}

\begin{theorem}\label{thm:3stgc}
For every $n\geq 1$, $A_n$ from Construction~\ref{const:skip3} is a complete $3$-SkTGC of length $n$. 
\end{theorem}

\begin{IEEEproof}
    It is immediate by construction that $\abs{A_{n+1}}=2\abs{A_n}$ for all $n\geq 1$, so the size of $A_n$ follows. 

    By inspection, $A_1$ and $A_2$ are both $1$-SkTGC (and therefore also $3$-SkTGC). For $n\geq 3$, we will prove by induction on $n$ that $A_n$ is a $3$-SkTGC, and that $A_n$ begins with 
    \begin{equation}
    \label{eq:3sktgcstart}
    \begin{bmatrix}
        0\dots  000\\
        0\dots  001\\
        0\dots  011\\
        \vdots\\
        1\dots  111
    \end{bmatrix},
    \end{equation}
    and ends with
    \begin{equation}
    \label{eq:3sktgcend}
    \begin{bmatrix}
    111 \dots  110\\
    011 \dots  110\\
    001 \dots  110\\
    \vdots \\
    000 \dots  010
    \end{bmatrix}.
    \end{equation}
    
    For the induction base, by inspection,
    \[A_3=\begin{bmatrix}000 \\ 001 \\ 011 \\ 111 \\ 101 \\ 100 \\ 110 \\ 010\end{bmatrix}\]
    is a $1$-SkTGC, and it satisfies~\eqref{eq:3sktgcstart} and~\eqref{eq:3sktgcend}.

    Now suppose that $A_n$ is a valid $3$-SkTGC, and that it satisfies the beginning and ending condition. By construction, $A_{n+1}$ begins with $\bzero$, and the following codewords are the same as that of $A_n$, but with a $1$ appended on the right, resulting in the desired starting sequence as~\eqref{eq:3sktgcstart}. The ending of $A_{n+1}$ is the beginning of $A_n$ reversed,  after removing the $\bzero$, and with a $0$ appended on the right, resulting in the desired ending sequence as~\eqref{eq:3sktgcend}.

    Given that $A_n$ is a valid $3$-SkTGC, we first observe that, by construction, the codewords of $A_{n+1}$ are all distinct. Additionally, the transitions inside each block are valid by the induction hypothesis. Thus, to see that $A_{n+1}$ is indeed a valid $3$-SkTGC, it only remains to check the transitions between the blocks. The transition between the second and first blocks is as follows: 
    \[
    \sparenv*{
    \begin{array}{c}
    0\dots  010\\
    0\dots  000\\
    \hline
    0\dots  001\\
    0\dots  011
    \end{array}
    }
    \]
    The changes are in positions $n, n+1, n$, so they are consecutive. If we list the last two codewords from the first block, and the first two in the second one, we can see that the transition between those blocks is as follows:
    \[
    \sparenv*{
    \begin{array}{c}
    0\dots  01101\\
    0\dots  00101\\
    \hline
    0\dots  00100\\
    0\dots  01100
    \end{array}
    }
    \]
    The changes are in positions $n-2, n+1, n-2$, which preserves the $3$-SkTGC condition. 
\end{IEEEproof}

Construction~\ref{const:skip3} allows straightforward decoding and encoding algorithms. However, since the these codes are only a stepping stone to obtaining complete $2$-SkTGCs and asymptotically non-vanishing $1$-SkTGCs, in the interest of brevity we omit these algorithms.

\subsection{The case $k=2$}

We now present a construction to obtain complete $2$-SkTGCs. The construction has the same flavor as Construction~\ref{const:skip3} for $3$-SkTGCs, but is more involved.

\begin{construction}\label{const:skip2}
Define the code
\[
A_3 \eqdef \begin{bmatrix} 000 \\ 001 \\ 011 \\ 111 \\ 110 \\ 100 \\ 101 \end{bmatrix}.
\]
Furthermore, for every $n\geq 3$, define the codes $A_{n+1}$, $B_{n+1}$ and $C_{n+1}$ as follows:
\begin{align*}
    B_{n+1} &\eqdef
    \sparenv*{
  \begin{array}{c|c}
    0\dots 0 & 0 \\
    \hline
     & 1 \\
    A_n & \vdots \\
     & 1 \\
     \hline
     & 0 \\
    \overleftarrow{A_n'} & \vdots \\
     & 0 
    \end{array}
    }
    \begin{array}{l}
    \text{ending of 2nd block} \\
    \phantom{1}\\
    \text{1st block}\phantom{\vdots}\\
    \phantom{1}\\
    \phantom{1}\\
    \text{2nd block}\phantom{\vdots}\\
    \phantom{1}\\
    \end{array} \\
    A_{n+1} &\eqdef 
    \begin{bmatrix}
        B_{n+1}^* \\
         \bfe_{n-1} \\
         \bfe_{n-1}+\bfe_{n+1}
    \end{bmatrix}
    =
    \begin{bmatrix}
        B_{n+1}^* \\
         0\dots 0100 \\
         0\dots 0101
    \end{bmatrix}
    ,\\
    C_{n+1} &\eqdef 
    \begin{bmatrix}
        \bzero \\
         \bfe_{n-1} \\
         \bfe_{n-1}+\bfe_{n+1} \\
        B_{n+1}' \\
    \end{bmatrix}
    =
    \begin{bmatrix}
         0\dots 0000 \\
         0\dots 0100 \\
         0\dots 0101 \\
        B_{n+1}' \\
    \end{bmatrix}
    .
\end{align*}
\end{construction}

\begin{theorem}\label{thm:2stgc}
For every $n\geq 4$, $C_n$ from Construction~\ref{const:skip2} is a complete $2$-SkTGC of length $n$.
\end{theorem}

In order to prove Theorem~\ref{thm:2stgc} we will first need the following lemma:

\begin{lemma}\label{lemma:2stgc}
    For every $n\geq 3$, all of the following hold, with regard to Construction~\ref{const:skip2}:
    \begin{itemize}
        \item $A_n$ is a non-cyclic $2$-SkTGC of size $2^n-1$ which does not contain the codeword $\bfe_{n-1}$;
        \item Both the first and the last changes in $A_n$ occur in position $n$;
        \item $B_{n+1}$ is a (cyclic) $2$-SkTGC of size $2^{n+1}-2$.
    \end{itemize}. 
\end{lemma}

\begin{IEEEproof}
    First notice that it is possible to prove by a straightforward inductive argument that $A_n$ and $B_n$ begin with ~\eqref{eq:3sktgcstart} and that $B_n$ ends with ~\eqref{eq:3sktgcend}.
        
    We prove the claims on $A_n$ by induction on $n$, and the claim on $B_{n+1}$ will follow from them. For the induction base, by inspection, $A_3$ is a non-cyclical $2$-SkTGC that does not contain $\bfe_{3-1}=\bfe_{2}$, and the first and last changes occur in the last position. 
    
    Now suppose $A_n$ satisfies the claims. We prove that $A_{n+1}$ does too. Since the first and the last changes in $A_n$ occur in position $n$, the changes in between the first and second  blocks of $B_{n+1}$ are $n,n+1,n$, and the changes in the end of the second block and the beginning of the first block are $n,n+1,n$. This, together with the fact that $A_n$ is a $2$-SkTGC and that by construction $B_{n+1}$ does not contain repeated codewords, implies that $B_{n+1}$ is a cyclic $2$-SkTGC.
    
    As already mentioned, $B_{n+1}$ ends in $\bfe_n$. To build $A_{n+1}$, we remove the last codeword in $B_{n+1}$. This implies that $A_{n+1}$ does not contain $\bfe_n$, as desired. 
    
    Since $A_n$ does not contain $\bfe_{n-1}$, by construction $B_{n+1}$ contains neither $\bfe_{n-1}$ nor $\bfe_{n-1}+\bfe_{n+1}$. This means that adding them at the end of $B_{n+1}^*$ does not  repeat codewords. Also, $B_{n+1}^*$ ends with
    \[
    \begin{bmatrix} 
    \bfe_n+\bfe_{n-1}+\bfe_{n-2} \\ \bfe_n+\bfe_{n-1}
    \end{bmatrix}
    =
    \begin{bmatrix}
    0\dots 01110 \\
    0\dots 00110
    \end{bmatrix}.
    \]
    Thus, $A_{n+1}$ ends with 
    \[
    \begin{bmatrix} 
    \bfe_n+\bfe_{n-1}+\bfe_{n-2} \\ \bfe_n+\bfe_{n-1} \\ \bfe_{n-1} \\ \bfe_{n-1}+\bfe_{n+1}
    \end{bmatrix}
    =
    \begin{bmatrix}
    0\dots 01110 \\
    0\dots 00110 \\
    0\dots 00100 \\
    0\dots 00101
    \end{bmatrix}.
    \]

    The last changes in $A_{n+1}$ are therefore in positions $n-2,n,n+1$. This shows that $A_{n+1}$ is a non-cyclic $2$-SkTGC, and the last change occurs in position $n+1$. Since $B_{n+1}^*$ begins with 
    \[
    \begin{bmatrix}
    0\dots 00\\
    0\dots 01
    \end{bmatrix},
    \]
    the first change in $A_{n+1}$ also occurs in position $n+1$. 

    Finally, let $a_n=\abs{A_n}$. By construction, $a_{n+1}= 2a_n+1$, and $a_3=7$. Solving this recursion yields $a_n= 2^n-1$. Since $\abs{B_{n+1}}=\abs{A_{n+1}}-1$, the claim on the size of $B_{n+1}$ follows as well.
\end{IEEEproof}

Now we can prove Theorem~\ref{thm:2stgc}.

\begin{IEEEproof}[Proof of Theorem~\ref{thm:2stgc}]
    By Lemma~\ref{lemma:2stgc}, for every $n\geq 4$ $B_n$ is a cyclic $2$-SkTGC  of size $2^n-2$, and since $A_{n-1}$ does not contain the codeword $\bfe_{n-2}$, this means that $B_n$ is missing precisely the codewords $\bfe_{n-2}$ and $\bfe_{n-2}+\bfe_n$. 

    Since the last, first, second and third codewords of $B_n$ are
    \[
    B_n = \begin{bmatrix}
        \bzero \\ \bfe_n \\ \bfe_n+\bfe_{n-1} \\ \vdots \\ \bfe_{n-1}
    \end{bmatrix} = 
    \begin{bmatrix}
        0\dots 000 \\ 0\dots 001 \\ 0\dots 011 \\ \vdots \\ 0\dots 010
    \end{bmatrix}
    \]
    we can insert the missing words between the first two codewords of $B_n$ to obtain $C_n$,
    \[
    C_n = \begin{bmatrix}
        \bzero \\ \bfe_{n-2} \\ \bfe_{n-2}+\bfe_n \\ \bfe_n \\ \bfe_{n-1}+\bfe_{n} \\ \vdots \\ \bfe_{n-1}
    \end{bmatrix} = 
    \begin{bmatrix}
        0\dots 0000 \\ 0\dots 0100 \\ 0\dots 0101 \\ 0\dots 0001 \\ 0\dots 0011 \\ \vdots \\ 0\dots 0010
    \end{bmatrix}
    \]
    so that the transition sequence (starting from the last codeword) is $n-1,n-2,n,n-2,n-1$, and the $2$-SkTGC property is preserved. This shows that $C_n$ is in fact a complete $2$-SkTGC.
\end{IEEEproof}

It is not hard to see that the induced transition graph for the codes $C_n$ as defined in Construction~\ref{const:skip2} is the following Toeplitz graph (and not a proper subgraph):

\begin{center}
    \begin{tikzpicture}
    \node (1) at (0,0) {1};
    \node (3) at (1,0) {3};
    \node (5) at (2,0) {5};
    \node (2) at (0.5,0.87) {2};
    \node (4) at (1.5,0.87) {4};
    \node (6) at (2.5,0.87) {6};
    \node (n-2) at (3.5,0.87) {$n-2$};
    \node (n-1) at (3,0) {$n-1$};
    \node (n) at (4,0) {$n$};
    \draw (1) -- (3);
    \draw (3) -- (5);
    \draw (2) -- (4);
    \draw (1) -- (2);
    \draw (2) -- (3);
    \draw (3) -- (4);
    \draw (4) -- (6);
    \draw (4) -- (5);
    \draw (5) -- (6);
    \draw (n-1) -- (n);
    \draw (n-2) -- (n);
    \draw (n-2) -- (n-1);
    \path (5) -- node[auto=false]{\ldots} (n-2);
\end{tikzpicture}
\end{center}
We comment that another, non-equivalent, complete binary $2$-SkTGC may be obtained as a side-effect of~\cite[Theorem 2.1]{WilErn02}, whose main purpose was to disprove a conjecture by Bultena and Ruskey. It is shown there that Gray codes may be constructed, whose transition graph is a tree of the following form:
\begin{center}
    \begin{tikzpicture}
    \node (5) at (0,0) {5};
    \node (3) at (0.7,0) {3};
    \node (1) at (1.4,0) {1};
    \node (7) at (1.4,0.8) {7};
    \node (9) at (2.1,0.8) {9};
    \node (2) at (2.1,0) {2};
    \node (4) at (2.8,0) {4};
    \node (11) at (2.8,0.8) {11};
    \node (6) at (3.5,0) {6};
    \node (13) at (3.5,0.8) {13};
    \node (2m-6) at (4.5,0) {$\scriptstyle 2m-6$};
    \node (2m-4) at (5.8,0) {$\scriptstyle 2m-4$};
    \node (2m-2) at (7.1,0) {$\scriptstyle 2m-2$};
    \node (2m) at (8.4,0) {$\scriptstyle 2m$};
    \node (2m+1) at (4.5,0.8) {$\scriptstyle 2m+1$};
    \draw (5) -- (3);
    \draw (3) -- (1);
    \draw (1) -- (7);
    \draw (1) -- (2);
    \draw (2) -- (9);
    \draw (2) -- (4);
    \draw (11) -- (4);
    \draw (6) -- (4);
    \draw (13) -- (6);
    \draw (2m+1) -- (2m-6);
    \draw (2m-4) -- (2m-6);
    \draw (2m-4) -- (2m-2);
    \draw (2m) -- (2m-2);
    \path (6) -- node[auto=false]{\ldots} (2m-6);
\end{tikzpicture}
\end{center}
The numbering of the nodes corresponds to the order in which the bit positions are added to the code, where in each step of the construction two bits are added. By permuting the code positions, we can obtain a complete $2$-SkTGC with the following transition graph:
\begin{center}
    \begin{tikzpicture}
    \node (1) at (0,0) {1};
    \node (2) at (0.7,0) {2};
    \node (3) at (1.4,0) {3};
    \node (4) at (1.4,0.8) {4};
    \node (6) at (2.1,0.8) {6};
    \node (5) at (2.1,0) {5};
    \node (7) at (2.8,0) {7};
    \node (8) at (2.8,0.8) {8};
    \node (9) at (3.5,0) {9};
    \node (10) at (3.5,0.8) {10};
    \node (2m-3) at (4.5,0) {$\scriptstyle 2m-3$};
    \node (2m-1) at (5.8,0) {$\scriptstyle 2m-1$};
    \node (2m) at (7.1,0) {$\scriptstyle 2m$};
    \node (2m+1) at (8.4,0) {$\scriptstyle 2m+1$};
    \node (2m-2) at (4.5,0.8) {$\scriptstyle 2m-2$};
    \draw (1) -- (2);
    \draw (2) -- (3);
    \draw (3) -- (4);
    \draw (3) -- (5);
    \draw (6) -- (5);
    \draw (7) -- (9);
    \draw (5) -- (7);
    \draw (7) -- (8);
    \draw (10) -- (9);
    \draw (2m+1) -- (2m);
    \draw (2m-1) -- (2m);
    \draw (2m-1) -- (2m-3);
    \draw (2m-2) -- (2m-3);
    \path (9) -- node[auto=false]{\ldots} (2m-3);
\end{tikzpicture}
\end{center}
However, the advantage of Construction~\ref{const:skip2} is that the ideas behind it can be adapted to obtain $1$-SkTGCs, as we will see later on, and it is more natural in the sense that bit columns are added in order and do not need to be permuted. 

We would now like to present decoding and encoding algorithms for Construction~\ref{const:skip2}. Since the difference between the codes $B_n$ and $C_n$ are minute, we opt for a cleaner presentation and present the algorithms for $B_n$. It is straightforward to adapt them for the complete codes $C_n$. Algorithm~\ref{alg:decode2} performs the decoding and Algorithm~\ref{alg:encode2} the encoding.

Given a codeword $\bfc$ in $B_n$, we can track its formation backwards in the construction. There are two options: either $\bfc$ comes from a codeword in $A_3$ to which $n-3$ bits were added on its right; or it comes from a codeword inserted in the code during some intermediate step $n_0$ (when building $A_{n_0}$ from $B_{n_0}$ we insert $\bfe_{n_0-2}$ and $\bfe_{n_0-2}+\bfe_{n_0}$). 

Thus, Algorithm~\ref{alg:decode2} checks for the existence of a step $n_0$ such that $\bfc[1\dots n_0]=\bfe_{n_0-2}$ or $\bfc[1\dots  n_0]=\bfe_{n_0-2}+\bfe_{n_0}$. Note that if such an $n_0$ exists, it is unique. It is easy to see that this can be done in linear time in $n$. If no such $n_0$ exists, the algorithm uses a lookup table to find the position of the first three bits of $\bfc$ in the base code $A_3$. After the first conditional statement, $\texttt{pos}$  contains the position of $\bfc[1\dots  n_0]$ in $A_{n_0}$.  

Notice that for $i\geq n_0+1$, the position of $\bfc[1\dots  i]$ in $A_i$ is actually the same as its position in $B_i$ (since $\bfc[1\dots  i]$ is not at the end of $B_i$). The final loop in the algorithm tracks the position of $\bfc[1\dots  i]$ in $A_i$ for $n_0+1\leq i\leq n$. If $\bfc[i]=1$, that means $\bfc[1\dots  i]$ is in the first block in the construction, and so its position is $1$ more than the position of $\bfc[1\dots  i-1]$ in $A_{i-1}$. If $\bfc[i]=0$, then $\bfc[1\dots  i]$ belongs to the second block. If the position of $\bfc[1\dots  i-1]$ was $0$, then the position of $\bfc[1\dots  i]$ is again $0$. In any other case, the position of $\bfc[1\dots  i]$ can be calculated as $2^{i-1}$ plus the distance of $\bfc[1\dots  i-1]$ to the ending position of $A_{i-1}$. This is $2^{i-1}+(2^{i-1}-2-\texttt{pos})=2^i-2-\texttt{pos}$, where $\texttt{pos}$ is the position of $\bfc[1\dots  i-1]$ in $A_{i-1}$. Notice that the expressions for case $\bfc[i]=0$ can be summarized as $(2^i-2-\texttt{pos})\bmod{2^i-2}$.

\begin{algorithm}[t]
\caption{Decoding algorithm for $B_n$ of Construction~\ref{const:skip2}}\label{alg:decode2}
\begin{algorithmic}
\State \textbf{Input:} $\bfc\in\Z_2^{n}$, $\bfc\neq \bfe_{n-2}, \bfe_{n-2}+\bfe_n$
\State \textbf{Output:} The index of $\bfc$ in $B_n$ of Construction~\ref{const:skip2}
\Function{Decode}{$\bfc,n$}

\State $\texttt{pos}\gets 0$
\State $n_0\gets 3$
\If{$\texttt{exists } 4\leq j\leq n-1 \texttt{ s.t. }$\\ \hspace{1cm}$\bfc[1\dots  j]=\bfe_{j-2}$}
    \State $n_0\gets j$
    \State $\texttt{pos}\gets 2^{n_0}-3$
    
\ElsIf{$\texttt{exists } 4\leq j\leq n-1 \texttt{ s.t. }$\\ \hspace{1cm} $\bfc[1\dots  j]=\bfe_{j}+\bfe_{j-2}$}
    \State $n_0\gets j$
    \State $\texttt{pos}\gets 2^{n_0}-2$
    
\Else
    \State $\texttt{pos}\gets \texttt{LookUp}_3[\bfc[1\dots  3]]$
\EndIf

\For{$\texttt{i in }[n_0+1\dots  n]$}
    \If{$\bfc[i]=0$}
        \State $\texttt{pos}\gets (2^{i}-2-\texttt{pos}) \bmod (2^i-2)$
    \Else
        \State $\texttt{pos}\gets 1+\texttt{pos}$
    \EndIf
\EndFor

\State \Return \texttt{pos}
\EndFunction
\end{algorithmic}
\end{algorithm}

Algorithm~\ref{alg:encode2} is an encoding algorithm for $B_n$, and is a simple reversal of the decoding algorithm.

\begin{algorithm}[t]
\caption{Encoding algorithm for $B_n$ of Construction~\ref{const:skip2}}\label{alg:encode2}
\begin{algorithmic}
\State \textbf{Input:} $n\geq 3$, $0\leq \texttt{pos}\leq 2^n-3$, 
\State \textbf{Output:} The codeword $\bfc$ with index $\texttt{pos}$ in $B_{n}$ in Construction~\ref{const:skip3} 

\Function{Encode}{$\texttt{pos},n$}
    \If{$n>3$}
    \If{$1\leq \texttt{pos}\leq 2^{n-1}-1$}
        \State $\bfc[n]\gets 1$
        \State $\texttt{pos}\gets \texttt{pos}-1$
    \Else
        \State $\bfc[n]\gets 0$
        \State $\texttt{pos}\gets (2^n-2-\texttt{pos})\bmod 2^n-2$
    \EndIf
    \For{$i \texttt{ in [n-1\dots  4]}$}
        \If{$\texttt{pos}=2^i-2$}
            \State $\bfc[1\dots  i] \gets \bfe_i+\bfe_{i-2}$
            \State \Return $\bfc$
        \ElsIf{$\texttt{pos}=2^i-3$}
            \State $\bfc[1\dots  i]\gets \bfe_{i-2}$
            \State\Return $\bfc$
        \ElsIf{$1\leq\texttt{pos}\leq 2^{i-1}-1$}
            \State $\bfc[i]\gets 1$
            \State $\texttt{pos}\gets \texttt{pos}-1$
        \Else
            \State $\bfc[i]\gets 0$
            \State $\texttt{pos}\gets (2^i-2-\texttt{pos})\bmod 2^i-2$
        \EndIf
    \EndFor
    \EndIf
    \State $\bfc[1\dots  3]\gets A_3[\texttt{pos}]$
    \State \Return $\bfc$
\EndFunction
\end{algorithmic}
\end{algorithm}

\subsection{The case $k=1$}

We finally tackle the most difficult case -- constructing $1$-SkTGCs. For simplicity of presentation, we shall allow positions to be numbered with indices in $\Z$. More precisely, in a code of length $N$, the positions will be numbered from left to right by 
\[-\ceil*{\frac{N-1}{2}},\dots  ,-1,0,1,\dots  ,\floor*{\frac{N-1}{2}},\]
and we denote 
\[n\eqdef\floor*{\frac{N-1}{2}}.\]

Before we elaborate the construction details, we would like to discuss the main strategy. The construction for $1$-SkTGCs presented in~\cite{WilBla06} yields codes with asymptotic rate $\frac{\log_2(3)}{2}\approx 0.792$. The main reason behind this is that the construction needs to discard a fourth of the code in each recursive step: in order to obtain a code of length $n+2$, the recursive construction requires a code of length $n$ such that the first change occurs in the last position, whilst the last change needs to occur in the first position. A straightforward reflecting construction is able to guarantee a change in the last position at the beginning, and a change in the first position only at around $3/4$ of the size of the code. Thus arises the need to discard the last quarter. 

Our Construction~\ref{const:skip1} overcomes this limitation by modifying the last codewords at each step so as to obtain a change in the first position. This imposes an extra requirement on the base code for the construction, and we will need to keep an invariant at each step that guarantees that the modification of the code will not repeat codewords. 

\begin{construction}\label{const:skip1}
Define the code $A_1$ as 
\[
A_1 \eqdef \begin{bmatrix} 
000 \\ 001 \\ 011 \\ 111
\end{bmatrix},
\]
and for every $n\geq 1$, define the codes $A_{n+1}$ and $B_{n+1}$ as follows:
\[
    B_{n+1} \eqdef \sparenv*{\begin{array}{c|c|c}
        0 & 0\dots  0 & 0 \\
        \hline
         0 &  & 1\\
        \vdots & A_n & \vdots \\
       0 &   & 1\\
         \hline
       1 &   & 1\\
      \vdots &   \overleftarrow{A_n} & \vdots \\
       1 &   & 1 \\
       \hline 
         1 &  & 0\\
        \vdots & A_n & \vdots \\
       1 &   & 0\\
         \hline
            0 &  & 0\\
        \vdots & \overleftarrow{A_n'} & \vdots\\
       0 &   & 0
    \end{array}}
    \begin{array}{l}
    \text{(ending of 4th block)} \\
    \phantom{1}\\
    \text{1st block}\phantom{\vdots}\\
    \phantom{1}\\
    \phantom{1}\\
    \text{2nd block}\phantom{\vdots}\\
    \phantom{1}\\
    \phantom{1}\\
    \text{3rd block}\phantom{\vdots}\\
    \phantom{1}\\
    \phantom{0}\\
    \text{4th block}\phantom{\vdots}\\
    \phantom{0}
    \end{array}
\]
and $A_{n+1}$ is obtained from $B_{n+1}$ by replacing the last ${n-1}$ codewords from $B_{n+1}$ by the $n+1$ codewords 
\[
\begin{bmatrix}
\bfe_{-1}+\sum_{i=1}^n \bfe_i\\
\bfe_{-2}+\bfe_{-1}+\sum_{i=1}^n \bfe_i\\
\vdots \\
\sum_{i=1}^{n+1}\bfe_{-i}+\sum_{i=1}^n \bfe_i
\end{bmatrix}
=
\sparenv*{
\begin{array}{c|c|c}
0\dots 001 & 0 & 1 \dots 10 \\
0\dots 011 & 0 & 1 \dots 10 \\
\vdots & \vdots & \vdots \\
1\dots 111 & 0 & 1 \dots 10 \\
\end{array}
},
\]
where the position with index $0$ is surrounded by vertical lines for emphasis.
\end{construction}

To prove that Construction~\ref{const:skip1} produces $1$-SkTGCs, we will need the following lemma:

\begin{lemma}\label{lemma:1SkTGC}
    For every $n\geq 1$, all of the following hold with regard to Construction~\ref{const:skip1}:
    \begin{itemize}
        \item $A_n$ is a non-cyclic $1$-SkTGC of length $2n+1$ which does not contain the codewords $\sum_{i=1}^j \bfe_{-i}+\sum_{i=1}^{n} \bfe_i$, for any $1\leq j\leq n$;
        \item The first change in $A_n$ occurs in position $n$, and the last, in position $-n$;
        \item $B_{n+1}$ is a (cyclic) $1$-SkTGC of length $2n+1$.
    \end{itemize}
\end{lemma}

\begin{IEEEproof}
As we have already done in previous proofs, we begin by observing that using a straightforward inductive argument proves that $A_n$ begins with
\begin{equation}
\label{eq:anbegin}
\begin{bmatrix}
\bzero \\
\bfe_n \\
\bfe_{n-1}+\bfe_n \\
\vdots\\
\sum_{i=0}^n \bfe_i
\end{bmatrix}
=
\sparenv*{
\begin{array}{c|c|c}
0\dots 0 & 0 & 0 \dots 00 \\
0\dots 0 & 0 & 0 \dots 01 \\
0\dots 0 & 0 & 0 \dots 11 \\
\vdots & \vdots & \vdots \\
0\dots 0 & 1 & 1 \dots 11 \\
\end{array}
},
\end{equation}
for all $n\geq 1$.

We prove the first two claims by induction on $n$, and the third claim will be proved as a side effect of that proof. For the induction base, by inspection, $A_1$ is a non-cyclic $1$-SkTGC, it does not contain the codeword $101=\bfe_1+\bfe_{-1}$, and the first and last changes occur in positions $1$ and $-1$ respectively. Now suppose $A_n$ satisfies the conditions, and we need to prove that $A_{n+1}$ does too. 

We contend that $B_{n+1}$ is a $1$-SkTGC. To prove that, we first observe that the codewords of $B_{n+1}$ are distinct since those of $A_n$ are by the induction hypothesis. Since $A_n$ is a non-cyclic $1$-SkTGC (again, by the induction hypothesis), we  only need to check the transitions between blocks to verify that $B_{n+1}$ is a $1$-SkTGC. The transitions between the fourth and first blocks, and between the second and third blocks, are valid since $A_n$ begins with a change in position $n$. The transitions between the first and the second blocks, and between the third and the fourth blocks, are valid because $A_n$ ends with a change in position $-n$. 

By~\eqref{eq:anbegin}, it follows that $B_{n+1}$ ends with 
\[
\sparenv*{\begin{array}{c}
\sum_{i=0}^n \bfe_i\\
\sum_{i=1}^n \bfe_i\\
\hline
\sum_{i=2}^n \bfe_i\\
\vdots\\
\bfe_{n-1}+\bfe_n \\
\bfe_n 
\end{array}
}
=
\sparenv*{
\begin{array}{c|c|c}
0\dots 0 & 1 & 11 \dots 110 \\
0\dots 0 & 0 & 11 \dots 110 \\
\hline
0\dots 0 & 0 & 01 \dots 110 \\
\vdots & \vdots & \vdots \\
0\dots 0 & 0 & 00 \dots 110 \\
0\dots 0 & 0 & 00 \dots 010 \\
\end{array}
},
\]
where the line denotes the location of last $n-1$ codewords. Thus, after removing the last $n-1$ codewords and adding the $n+1$ words specified by the construction, $A_{n+1}$ ends with
\[
\sparenv*{\begin{array}{c}
\sum_{i=0}^n \bfe_i\\
\sum_{i=1}^n \bfe_i\\
\hline
\bfe_{-1}+\sum_{i=1}^n \bfe_i\\
\bfe_{-2}+\bfe_{-1}+\sum_{i=1}^n \bfe_i\\
\vdots\\
\sum_{i=1}^{n+1}\bfe_{-i}+\sum_{i=1}^n \bfe_i\\
\end{array}
}
=
\sparenv*{
\begin{array}{c|c|c}
0\dots 000 & 1 & 1 \dots 10 \\
0\dots 000 & 0 & 1 \dots 10 \\
\hline
0\dots 001 & 0 & 1 \dots 10 \\
0\dots 011 & 0 & 1 \dots 10 \\
\vdots & \vdots & \vdots \\
1\dots 111 & 0 & 1 \dots 10 \\
\end{array}
}.
\]
so the last changes in $A_{n+1}$ are in positions $0,-1,\dots , {-(n+1)}$. In particular, the $1$-SkTGC property is preserved, and $A_{n+1}$ ends with a change in position $-(n+1)$. We also note that since $A_{n+1}$ starts with $\bzero$ followed by $\bfe_{n+1}$, the first change is in position $n+1$. 
    
We need to verify that $A_{n+1}$ does not repeat codewords. Since by the induction hypothesis $A_n$ does not contain $\sum_{i=1}^j \bfe_{-i}+\sum_{i=1}^n \bfe_i$ for any $1\leq j\leq n$, then $B_{n+1}$ does not contain these codewords either, nor does it contain $\sum_{i=1}^{n+1} \bfe_{-i}+\sum_{i=1}^n \bfe_i$. This means that the newly added words in $A_{n+1}$ were not present in $B_{n+1}$, and thus are not repeated. 

Finally, $A_{n+1}$ does not contain any word $\sum_{i=1}^j \bfe_{-i}+\sum_{i=1}^{n+1}\bfe_i$, for $1\leq j\leq n+1$, since all of the added words have a $0$ in position $n+1$, and since $A_n$ does not contain $\sum_{i=1}^j \bfe_{-i}+\sum_{i=1}^n \bfe_i$ for any $1\leq j\leq n$. This concludes the induction step. 
\end{IEEEproof}

\begin{theorem}\label{thm:1SkTGC}
    For every $n\geq 2$, $B_n$ as defined in Construction~\ref{const:skip1} is a $1$-SkTGC of length $N=2n+1$ and size $\frac{7}{12}2^{N}-\frac{8}{3}$.
\end{theorem}

\begin{IEEEproof}
    By Lemma~\ref{lemma:1SkTGC}, $B_n$ is a $1$-SkTGC for every $n\geq 2$. As for the size of the code, denote $a_n=\abs{A_n}$ and $b_n=\abs{B_n}$. Then, $a_1=4$, $b_{n+1}=4a_n$, and $a_n=b_n+2$. We can set $b_1=2$, so we obtain the following recursion:
    \[
        \begin{cases}
            b_1=2\\
            b_{n+1}=4b_n+8 & \text{for $n\geq 2$}
        \end{cases}
    \]
    Solving the recursion yields $b_n=\frac{7}{12}2^{2n+1}-\frac{8}{3}$.
\end{IEEEproof}

Construction~\ref{const:skip1} can be easily adapted to obtain codes of even length:

\begin{construction}
\label{const:skip1even}
Define the code $A_1$ as
\[
A_1 \eqdef \begin{bmatrix} 0000 \\ 0001 \\ 0011 \\ 0111 \\ 1111 \end{bmatrix},
\]
and for every $n\geq 1$, $B_{n+1}$ is defined in the same way as in Construction~\ref{const:skip1}, and $A_{n+1}$ is obtained from $B_{n+1}$ by replacing the last $n-1$ codewords from $B_{n+1}$ by the $n+2$ codewords
\[
\begin{bmatrix}
\bfe_{-1}+\sum_{i=1}^n \bfe_i\\
\bfe_{-2}+\bfe_{-1}+\sum_{i=1}^n \bfe_i\\
\vdots \\
\sum_{i=1}^{n+2}\bfe_{-i}+\sum_{i=1}^n \bfe_i
\end{bmatrix}
=
\sparenv*{
\begin{array}{c|c|c}
0\dots 001 & 0 & 1 \dots 10 \\
0\dots 011 & 0 & 1 \dots 10 \\
\vdots & \vdots & \vdots \\
1\dots 111 & 0 & 1 \dots 10 \\
\end{array}
},
\]
where the position with index $0$ is surrounded by vertical lines for emphasis.
\end{construction}

\begin{theorem}\label{thm:1sktgceven}
For every $n\geq 2$, $B_n$ as defined in Construction~\ref{const:skip1even} is a $1$-SkTGC of length $N=2n+2$ and size $\frac{3}{8}2^{N}-4$.
\end{theorem}
\begin{IEEEproof}
The proof is the same as that of Theorem~\ref{thm:1SkTGC}.
\end{IEEEproof}

Let us now consider decoding and encoding algorithms for the odd-length $1$-SkTGC from Construction~\ref{const:skip1} (the algorithms for the even-length $1$-SkTGC from Construction~\ref{const:skip1even} are similar). We start with Algorithm~\ref{alg:decode1} for decoding $B_n$. Since the code is not complete, the algorithm may return an error if presented with an input that is not a codeword. The logic behind the algorithm is quite similar to that of Algorithm~\ref{alg:decode2}. 
First, let us denote
\[L(n)\eqdef\abs*{A_n}=\frac{7}{12}2^{2n+1}-\frac{2}{3}.\]

Given a codeword $\bfc$ in $B_n$, we can trace back its formation. It can either come from a word in $A_1$, or from a word of the form 
\begin{equation}
\label{eq:form}
\sum_{i=1}^{n_0-1}\bfe_i+\sum_{i=1}^j \bfe_{-i},
\end{equation}
for some step $n_0\leq n-1$, and $1\leq j\leq n_0$, which was added when constructing $A_{n_0}$ from $B_{n_0}$. The first conditional statement finds an adequate $n_0$ if it exists, and calculates the position of $\bfc[-n_0\dots  n_0]$ in $A_{n_0}$. It is not hard to see that this can be done in linear time. Notice that we need to handle the case in which $\bfc$ does not belong to the code. This could happen for one of the two following reasons: either $\bfc$ is a codeword which was removed when obtaining $A_m$ from $B_m$, for $2\leq m\leq n-1$, in which case $\bfc[-m\dots  m]=\sum_{i=j}^{m-1}\bfe_i$ for some $j\geq 2$, or $\bfc$ is a codeword which is not of the form~\eqref{eq:form} and $\bfc[-1\dots  1]\not\in A_1$. We use a lookup table $\texttt{LookUp}_1$ that contains the indexes for all codewords in $A_1$.

We index the codewords of $B_n$ using $L(n)$ in the following way:
\[
B_n=
    \sparenv*{\begin{array}{c|c|c}
        0 & 0\dots  0 & 0 \\
        \hline
         0 &  & 1\\
        \vdots & A_{n-1} & \vdots \\
       0 &   & 1\\
         \hline
       1 &   & 1\\
      \vdots &   \overleftarrow{A_{n-1}} & \vdots \\
       1 &   & 1 \\
       \hline 
         1 &  & 0\\
        \vdots & A_{n-1} & \vdots \\
       1 &   & 0\\
         \hline
            0 &  & 0\\
        \vdots & \overleftarrow{A_{n-1}'} & \vdots\\
       0 &   & 0
    \end{array}}
    \begin{array}{l}
    0 \\
    1\\
    \vdots\\
    L(n-1)\\
    L(n-1)+1\\
    \vdots\\
    2L(n-1)\\
    2L(n-1)+1\\
    \vdots\\
    3L(n-1)\\
    3L(n-1)+1\\
    \vdots\\
    4L(n-1)-1
    \end{array}
\]
With this indexing in mind, we can deduce how to obtain the position of $\bfc[-(i+1)\dots  i+1]$ in $B_{i+1}$ (which is the same as its position in $A_{i+1}$ when $i\geq n_0$) from the position of $\bfc[-i\dots  i]$ in $A_i$. 

\begin{algorithm}[t]
\caption{Decoding algorithm for Construction~\ref{const:skip1}}\label{alg:decode1}
\begin{algorithmic}
\State \textbf{Input:} $n\geq 2$, $\bfc\in\mathbb{Z}_2^{2n+1}$ 
\State \textbf{Output:} The index of $\bfc$ in $B_n$ from Construction~\ref{const:skip1}, or an error if $\bfc\not\in B_n$
\Function{Decode}{$\bfc,n$}
\State $n_0\gets 1$
\If{${\texttt{exists } 2\leq m\leq n-1 \texttt{ and } 1\leq j\leq m \texttt{ s.t. }}$\\ \hspace{1cm}$\bfc[-m\dots  m]=\sum_{i=1}^{m-1}\bfe_i+\sum_{i=1}^{j}\bfe_{-i}$}
    \State $n_0\gets m$
    \State $\texttt{pos}\gets L(n_0)-n_0-1+j$
    
\ElsIf{${\texttt{exists } 2\leq m\leq n-1 \texttt{ and } 2\leq j\leq m-1}$ \\ \hspace{1cm}$ \texttt{ s.t. }\bfc[-m\dots  m]=\sum_{i=j}^{m-1}\bfe_i$}
    \State \Return \texttt{Error}
\ElsIf{$\bfc[-1\dots  1]\in A_1$}
    \State $\texttt{pos} \gets \texttt{LookUp}_1[\bfc[-1\dots  1]]$
    \State $n_0\gets 1$
\Else
    \State \Return \texttt{Error}
\EndIf

\For{$\texttt{i in }[n_0+1\dots  n]$}
    \If{$\bfc[i]\neq 0 \texttt{ and } \bfc[-i]=0$}
        \State $\texttt{pos}\gets 1+\texttt{pos}$
    \ElsIf{$\bfc[i]\neq 0 \texttt{ and } \bfc[-i]\neq 0$}
        \State $\texttt{pos}\gets 2L(i-1)-\texttt{pos}$
    \ElsIf{$\bfc[i]= 0 \texttt{ and } \bfc[-i]\neq 0$}
        \State $\texttt{pos}\gets 2L(i-1)+1+\texttt{pos}$
    \Else
        \State $\texttt{pos}\gets (4L(i-1)-\texttt{pos}) \bmod 4L(i-1)$
    \EndIf
\EndFor
\State \Return $\texttt{pos}$
\EndFunction
\end{algorithmic}
\end{algorithm}

Encoding can also be performed in linear time, as shown in Algorithm~\ref{alg:encode1}. For simplicity, we omit certain details, such as the explicit calculations to determine to which block in $B_i$ a certain position belongs. We also denote the code 
\[\begin{bmatrix} 00 \\ 01 \\ 11 \\ 10\end{bmatrix}\]
by $\texttt{Gray}_2$, and we use the function $\texttt{ConvertPos(pos,i)}$, which given a position in $B_i$, returns the position it corresponds to in $A_{i-1}$.

\begin{algorithm}[t]
\caption{Encoding algorithm for Construction~\ref{const:skip1}}\label{alg:encode1}
\begin{algorithmic}
\State \textbf{Input:} $n\geq 2$, $0\leq \texttt{pos}\leq b_n-1$
\State \textbf{Output:} The codeword $\bfc$ with index $\texttt{pos}$ in $B_n$ from Construction~\ref{const:skip1}
\Function{Encode}{$\texttt{pos},n$}
        \If{$\texttt{pos}\in \texttt{Block}_i \texttt{ of }B_n$}
            \State $\bfc[-n,n]\gets \texttt{Gray}_2[i \bmod 4]$
            \State $\texttt{pos}\gets \texttt{ConvertPos(pos,n)}$
        \EndIf
        \For{$\texttt{i in }[n-1\dots  2]$}
            \If{$\texttt{pos}\geq L(i)-i$}
                \State $\bfc[-i\dots  i]\gets \sum_{j=1}^{i-1}\bfe_j+\sum_{j=1}^{\texttt{pos}-L(i)+i+1}\bfe_{-j}$
                \State \Return $\bfc$
            \ElsIf{$\texttt{pos}\in \texttt{Block}_j \texttt{ of }B_i$}
                \State $\bfc[-i,i]\gets \texttt{Gray}_2[j \bmod 4]$
                \State $\texttt{pos}\gets \texttt{ConvertPos(pos,i)}$
            \EndIf
        \EndFor
    \State $\bfc[-1..1]\gets A_1[\texttt{pos}]$
    \State \Return $\bfc$
\EndFunction
\end{algorithmic}
\end{algorithm}

We conclude this section by suggesting a method for optimizing the codes from Constructions~\ref{const:skip1} and~\ref{const:skip1even}. Theorems~\ref{thm:1SkTGC} and~\ref{thm:1sktgceven} show that it is possible to obtain $1$-SkTGCs whose size is asymptotically a constant fraction of a complete code, i.e., size that is $\approx c\cdot 2^n$ for some real constant $c$. Interestingly, the constant for Construction~\ref{const:skip1} is $\frac{7}{12}$, which differs from the constant $\frac{3}{8}$ of Construction~\ref{const:skip1even}. The constants obtained in the previous section are not best possible: we can improve them by choosing larger base cases and by using a slightly generalized version of the construction, which we now describe.

Suppose we number the positions of codewords of length  $N$ with indices $-L,\dots  ,-1,0,1,\dots  ,R$, where $L\geq 1$, $R\geq 1$, and $N=L+R+1$ (so the numbering is not necessarily symmetric around $0$). 

\begin{construction}\label{const:skip1general}
    Let $A_0$ be a non-cyclic $1$-SkTGC of length $N_0=L+R+1$, $L,R\geq 1$, that satisfies the following conditions:
    \begin{itemize}
        \item The first change occurs in the last bit (bit $R$).
        \item The last change occurs in the first bit (bit $-L$).
        \item It starts with 
\[
\begin{bmatrix}
\bzero\\
\bfe_R \\
\bfe_{R-1}+\bfe_R\\
\vdots\\
\sum_{i=0}^R \bfe_i
\end{bmatrix}
=
\sparenv*{
\begin{array}{c|c|c}
0\dots 0 & 0 & 0 \dots 000 \\
0\dots 0 & 0 & 0 \dots 001 \\
0\dots 0 & 0 & 0 \dots 011 \\
\vdots & \vdots & \vdots \\
0\dots 0 & 1 & 1 \dots 111 \\
\end{array}
}.
\]
        \item It does not contain any of the codewords $\sum_{i=1}^j \bfe_{-i}+\sum_{i=1}^{R} \bfe_i$ for $1\leq j\leq L$. 
    \end{itemize}
    For every $n\geq 0$, $B_{n+1}$ is defined in the same way as in Construction~\ref{const:skip1}, and $A_{n+1}$ is obtained from $B_{n+1}$ by replacing the last $R+n-1$ codewords with the $L+n+1$ codewords
\[
\begin{bmatrix}
\bfe_{-1}+\sum_{i=1}^{R+n} \bfe_i\\
\bfe_{-2}+\bfe_{-1}+\sum_{i=1}^{R+n} \bfe_i\\
\vdots \\
\sum_{i=1}^{L+n+1}\bfe_{-i}+\sum_{i=1}^n \bfe_i
\end{bmatrix}
=
\sparenv*{
\begin{array}{c|c|c}
0\dots 001 & 0 & 1 \dots 10 \\
0\dots 011 & 0 & 1 \dots 10 \\
\vdots & \vdots & \vdots \\
1\dots 111 & 0 & 1 \dots 10 \\
\end{array}
},
\]
where the position with index $0$ is surrounded by vertical lines for emphasis.
\end{construction}

Notice that Constructions~\ref{const:skip1} and~\ref{const:skip1even} are both particular cases of Construction~\ref{const:skip1general}, where $L=1$, $R=1$, and $L=2$, $R=1$, respectively. 

\begin{theorem}\label{thm:skip1general}
    For every $n\geq 1$, $B_n$ as defined in Construction~\ref{const:skip1general} is a $1$-SkTGC of length $N=L+R+1+2n$ and size 
    \[
         2^{N}\parenv*{\frac{\abs{A_0}+\frac{1}{3}(L-R+2)}{2^{N_0}}}-\frac{4}{3}(L-R+2).  
    \]
\end{theorem}

\begin{IEEEproof}
The proof that $B_n$ is a $1$-SkTGC is essentially the same as that of Theorem~\ref{thm:1SkTGC}. For the size of the code, denote $a_i\eqdef\abs{A_i}$ and $b_i\eqdef\abs{B_i}$. We know that for $n\geq 1$, 
\[b_{n+1}=4a_n=4(b_n-(R+n-2)+(L+n))=4(b_n+L-R+2).\]
Notice that we may extend the definition of $b_n$ to $n\geq 0$ by defining $b_0\eqdef a_0-(L-R+2)$. Solving the recursion yields the claimed size.
\end{IEEEproof}

We would like to find base cases such that the multiplicative constant in Theorem~\ref{thm:skip1general} is as large as possible. Notice that the difficulty of doing this is comparable (if not harder) to that of the original problem we are trying to solve: we need to find large $1$-SkTGCs which satisfy additional conditions as specified in Construction~\ref{const:skip1general}. 

Table~\ref{table:constants} shows the best possible constants for different combinations of the parameters in Construction~\ref{const:skip1general}, obtained by an exhaustive computer search. Highlighted in gray are the best constants found for constructions of even and odd length. In the Appendix we include base cases that attain these two values. 

\begin{table}[t]
\caption{Best constants for Construction~\ref{const:skip1general}}
\centering
  \begin{tabular}{ccccl}
    \hline
    $N_0$ & $L$ & $R$ & $a_0$ & $c$ \\
    \hline\hline
    4 & 2 & 1 & 11 & 0.75  \\
    4 & 1 & 2 & 9 & $0.58\overline{3}$\\
    \hline
    5 & 3 & 1 & 18 & $0.6041\overline{6}$ \\
    5 & 2 & 2 & 18 & $0.58\overline{3}$ \\
    5 & 1 & 3 & 24 & 0.75\\
    \hline
    \cellcolor{gray!10}6 & 4 & 1 & 47 & \cellcolor{gray!10}$0.76041\overline{6}$\\
    6 & 3 & 2 & 47 & 0.75 \\
    6 & 2 & 3 & 45 & $0.708\overline{3}$\\
    6 & 1 & 4 & 47 & $0.7291\overline{6}$\\

    \hline
    7 & 5 & 1 & 104 & $0.828125$\\
    7 & 4 & 2 & 104 & $0.82291\overline{6}$\\
    \cellcolor{gray!10}7 & 3 & 3 & 108 & \cellcolor{gray!10}$0.848958\overline{3}$\\
    7 & 2 & 4 & 108 & $0.84375$\\
    7 & 1 & 5 & 104 & $0.807291\overline{6}$\\
    \hline
    
\end{tabular}  
  \label{table:constants}
\end{table}

\section{ $m$-ary skew-tolerant Gray codes}
\label{sec:nonbinary}

In this section we construct non-binary $1$-SkTGCs. Unlike the previous section, all of the constructed codes are complete. We will divide our constructions into three cases: $m\geq 5$, which is easier to handle as the higher number of symbols allows us to have many consecutive changes in the same position, and this can be exploited to obtain a straightforward construction; $m=4$, for which we will use binary $2$-SkTGCs; and finally $m=3$, for which we first develop a construction that yields non-cyclic codes, and then derive cyclic codes.

\subsection{$m$-ary codes, $m\geq 5$} 
We start with the simplest case, where the alphabet $\Z_m$, is large enough, namely, $m\geq 5$.
\begin{construction}\label{const:base5}
Define the code
\[
A_1\eqdef \begin{bmatrix} 0 \\ 1 \\ \vdots \\ m-1 \end{bmatrix},
\]
and for every $n\geq 1$ define $A_{n+1}$ as
\[
A_{n+1}\eqdef \sparenv*{
\begin{array}{c|c}
\bfc_n & m-3 \\
\bfc_n & m-4 \\
\bfc_n & m-5 \\
\vdots & \vdots \\
\bfc_n & 0 \\
A'_n & 0 \\
\overleftarrow{A_n'} & 1 \\
\vdots & \vdots \\
A'_n \text{ or } \overleftarrow{A_n'} & m-1 \\
\bfc_n & m-1 \\
\bfc_n & m-2
\end{array}
}
\]
where $\bfc_n$ is the first codeword of $A_n$, and the entry ``$A'_n \text{ or } \overleftarrow{A_n'}$'' is $A_n'$ if $m$ is odd, and $\overleftarrow{A_n'}$ if $m$ is even.
\end{construction}

\begin{theorem}\label{thm:base5}
    For every $n\geq 1$, and every $m\geq 5$, $A_n$ as defined in Construction~\ref{const:base5} is a complete $m$-ary $1$-SkTGC. 
\end{theorem}

\begin{IEEEproof}
We will prove by induction that $A_n$ is a complete $m$-ary $1$-SkTGC such that the first two changes occur in  position $n$, as well as the last change, and the change between the ending and the beginning of $A_n$. For the induction basis, this holds for $A_1$ by inspection. 

Assume this holds for $A_n$, and we prove it also holds for $A_{n+1}$. Using the induction hypothesis, we immediately see that $A_{n+1}$ contains distinct codewords, and that
\[
\abs*{A_{n+1}}=m\abs*{A_n}=m^{n+1},
\]
and so $A_{n+1}$ is a complete code. We only need to check that the transitions between blocks of $A_{n+1}$ are valid. This follows from:
\begin{itemize}
\item $A_n$ begins with two changes in  position $n$, so $A_n'$ begins with a change in position $n$.
\item $A_n$ ends with a change in position $n$.
\item $\bfc_n$ is the first word of $A_n$, so from $\bfc_n$ to $A_n'$ there is a change in position $n$, and the same applies to the transition between $\overleftarrow{A_n'}$ and $\bfc_n$, and the transition between $A_n'$ and $\bfc_n$ (since by the induction hypothesis the change between the ending of $A_n$ and the beginning occurs in position $n$).
\end{itemize}  

From the construction, we can see that $A_{n+1}$ ends with a change in its last position, it begins with two changes in the last position, and the change between the last codeword and the first occurs in the last position as well. Finally, by inspection and the induction hypothesis, all of the transitions are of the form $\pm\bfe_i$ for some $i$. This completes the proof.
\end{IEEEproof}

\subsection{Quaternary codes}

Construction~\ref{const:base5} fails for $m<5$. In this section we present a simple construction for complete $1$-SkTGCs when $m=4$, which is based on Construction~\ref{const:skip2} of complete binary $2$-SkTGCs.

Define the following mapping, $\varphi:\Z_2^2\to \Z_4$,
    \[
    \varphi(00)=0, \quad
    \varphi(01)=1,\quad
    \varphi(11)=2, \quad
    \varphi(10)=3.
    \]
If $\bfc\in\Z_2^{2n}$ is a vector of even length, $\bfc=(c_1,\dots,c_{2n})$, we use $\varphi(\bfc)$ to denote the application of $\varphi$ to the (non-overlapping) pairs of bits of $\bfc$, namely,
\[
\varphi(\bfc)\eqdef( \varphi(c_1,c_2),\varphi(c_3,c_4),\dots,\varphi(c_{2n-1},c_{2n})).
\]
By natural extension, if $C=\bfc_0,\dots,\bfc_{P-1}$ is a sequence of vectors of even length, we define
\[ \varphi(C) \eqdef \varphi(\bfc_0),\dots,\varphi(\bfc_{P-1}).\]

\begin{construction}\label{const:base4}
    Given $n\geq 1$, Let $B_n$ be a complete binary $2$-SkTGC of length $2n$ (for example, $C_{2n}$ as defined in Construction~\ref{const:skip2} when $n\geq 2$; and the complete $1$-SkTGC of length $2$ for $n=1$). Construct the quaternary code
    \[ A_n = \varphi(B_n).\]
\end{construction}

\begin{theorem}\label{thm:base4}
    For every $n\geq 1$, $A_n$ as defined in Construction~\ref{const:base4} is a complete quaternary $1$-SkTGC.
\end{theorem}

\begin{IEEEproof}
The code $B_n$ is a complete Gray code of length $2n$, so it has $2^{2n}$ distinct codewords. Since $\varphi$ is invertible, the codewords of $A_n$ are also distinct and
\[ \abs*{A_n}=\abs*{B_n}=2^{2n}=4^n,\]
hence $A_n$ is complete.

The code $B_n$ is $2$-SkTGC, hence, any two consecutive changes in it are at most at distance $2$ apart. Thus, after applying $\varphi$, consecutive changes can either occur in the same position, or in adjacent positions. Finally, notice that if $\bfx,\bfy\in\Z_2^2$ are two-bit words which differ in only one bit, then 
\[\varphi(\bfx)-\varphi(\bfy)\equiv \pm 1 \pmod{4}.\]
This implies that for any two adjacent codewords in $A_n$, $\bfc_i$ and $\bfc_{i+1}$,
\[\bfc_{i+1}-\bfc_{i} = \pm\bfe_{\delta_i},\]
where indices are taken modulo $4^n$.
\end{IEEEproof}

\subsection{Ternary codes}

We first describe a procedure to obtain ternary non-cyclic $1$-SkTGCs, and then we use them to derive cyclic codes.

\begin{construction}
\label{const:base3noncyc}
Define the code $A_2$ as:
\[
A_2 \eqdef \begin{bmatrix} 00 \\ 01 \\ 02 \\ 22 \\ 20 \\ 21 \\ 11 \\ 12 \\ 10 \end{bmatrix}.
\]
For every $n\geq 2$ we define $A_{n+1}$ as
\[
A_{n+1}\eqdef \sparenv*{
\begin{array}{c|c}
\bfc_n & 0 \\
\bfc_n & 2 \\
\bfc_n & 1 \\
\widehat{A_n}   &  1\\
\overleftarrow{\widehat{A_n}}   &  0\\
\widehat{A_n} & 2 \\
\bfd_n & 2 \\
\bfd_n & 1 \\
\bfd_n & 0
\end{array}
},
\]
where $\bfc_n$ and $\bfd_n$ are the first and last words of $A_n$, respectively. 
\end{construction}

\begin{theorem}
\label{thm:base3noncyclic}
For every $n\geq 2$, $A_n$ as defined in Construction~\ref{const:base3noncyc} is a complete ternary $1$-SkTGC which begins with $\bzero$ and ends with $\bfe_1$. For $n\geq 3$, $A_n$ is non-cyclic.
\end{theorem}

\begin{IEEEproof}
By construction, $A_n$ begins and ends with two changes in the $n$-th position. Thus, $\widehat{A_n}$ begins and ends with a change in the $n$-th position. This guarantees that the transitions between blocks in the construction are valid. It is immediate by construction that the codes do not repeat words, and that they contain all possible words of the corresponding length.

Note that $A_2$ begins with $00$ and ends with $10$, and in each step, $A_{n+1}$ begins and ends with the same words as $A_n$, with a $0$ on the right. This proves the statement about the beginning and ending of $A_n$. Finally, by simple induction $A_n$ is $1$-SkTGC. 
\end{IEEEproof}

\begin{construction}
\label{const:base3cyclic}
For all $i$, let $A_i$ be the code from Construction~\ref{const:base3noncyc}. For each $n\geq 3$ define the code $B_{n}$ of length $n$ as
\[
    B_n \eqdef \sparenv*{
    \begin{array}{c|c}
       A_{n-1}  &  0\\
       \bfe_1+A_{n-1}  &  1\\
       2\bfe_1+A_{n-1}  &  2
    \end{array}
    }
\]
where $\bfe_1+A_{n-1}$ is the code obtained by adding $\bfe_1$ to each codeword of $A_{n-1}$, and similarly for $2\bfe_1+A_{n-1}$.
\end{construction}

\begin{theorem}\label{thm:base3cyclic}
For every $n\geq 3$, $B_n$ as defined in Construction~\ref{const:base3cyclic} is a complete ternary $1$-SkTGC of length $n$.
\end{theorem}
\begin{IEEEproof}
By Theorem~\ref{thm:base3noncyclic}, $A_{n-1}$ is a complete (non-cyclic) ternary $1$-SkTGC. Thus, the code $e_1+A_{n-1}$ satisfies this same property, and so does $2e_1+A_{n-1}$. It follows that all the codewords of $B_n$ are distinct, and that $\abs{B_n}=3^n$, hence it is complete. Within each block the code is SkTGC, so it only remains to verify the transitions between the blocks.

Again by Theorem~\ref{thm:base3noncyclic}, for every $n\geq 3$ the last codeword of $A_{n-1}$ is $\bfe_1$, which is the same as the first word of $\bfe_1+A_{n-1}$. The last word of $\bfe_1+A_{n-1}$ is $\bfe_1+\bfe_1=2\bfe_1$, which is the same as the first of $2\bfe_2+A_{n-1}$. The last word of $2\bfe_2+A_{n-1}$ is $2\bfe_1+\bfe_1=\bzero$, which is the first of $A_{n-1}$. Also, $A_{n-1}$, $\bfe_1+A_{n-1}$ and $2\bfe_1+A_{n-1}$ all begin and end with a change in the $(n-1)$-st position. This ensures that the block transitions are valid.
\end{IEEEproof}

As a concluding remark, notice that Constructions~\ref{const:base3noncyc} and~\ref{const:base3cyclic} can be generalized for every odd alphabet size $m\geq 3$, yielding an alternative to Construction~\ref{const:base5} when $m$ is odd.

\section{Conclusion}

In this paper we presented a construction for binary SkTGCs that asymptotically attains a constant fraction of possible codewords. This significantly improves upon the size of the best known construction. We also showed that it is possible to obtain complete codes by slightly relaxing the skew-tolerance condition and allowing consecutive changes to occur at a distance at most $2$. All these constructions have linear-time encoding and decoding algorithms. We also constructed complete non-binary SkTGCs.

There still remain many interesting questions. First and foremost, Slater's conjecture on the nonexistence of $1$-SkTGCs is still open: we do not know how to show even one word must be missed when constructing SkTGCs. We also do not know if it is possible to improve the size of the constructed codes and obtain asymptotically complete codes (that is, the limit of the ratio between the size of the code of length $n$ and $2^n$ converges to $1$). A possible strategy to do so could be to try to imitate Construction~\ref{const:skip2} of $2$-SkTGCs, in which the missing codewords from previous steps are inserted into the code later on. However, allowing jumps of size one imposes a huge limitation and it does not seem clear how to proceed in this case. 

In the absence of any known construction for complete binary $1$-SkTGCs (apart for three small values of $n$), another open question deals with complete binary $2$-SkTGCs. Can we construct complete binary $2$-SkTGCs whose transition sequence $\delta_0,\dots,\delta_{2^n-1}$ contains the least amount of pairs of consecutive entries differing by a value of $2$? Construction \ref{const:skip2} and the construction in \cite{WilErn02} are based on reflections, so in each step the number of jumps of size $2$ at least duplicates, and thus this quantity is exponential in $n$. Reducing this number will provide the code with better protection against skew errors.

\bibliographystyle{IEEEtranS}
\bibliography{allbib}

% Generated by IEEEtranS.bst, version: 1.14 (2015/08/26)
\begin{thebibliography}{10}
\providecommand{\url}[1]{#1}
\csname url@samestyle\endcsname
\providecommand{\newblock}{\relax}
\providecommand{\bibinfo}[2]{#2}
\providecommand{\BIBentrySTDinterwordspacing}{\spaceskip=0pt\relax}
\providecommand{\BIBentryALTinterwordstretchfactor}{4}
\providecommand{\BIBentryALTinterwordspacing}{\spaceskip=\fontdimen2\font plus
\BIBentryALTinterwordstretchfactor\fontdimen3\font minus \fontdimen4\font\relax}
\providecommand{\BIBforeignlanguage}[2]{{%
\expandafter\ifx\csname l@#1\endcsname\relax
\typeout{** WARNING: IEEEtranS.bst: No hyphenation pattern has been}%
\typeout{** loaded for the language `#1'. Using the pattern for}%
\typeout{** the default language instead.}%
\else
\language=\csname l@#1\endcsname
\fi
#2}}
\providecommand{\BIBdecl}{\relax}
\BIBdecl

\bibitem{AbbKat91}
H.~L. Abbot and M.~Katchalski, ``On the construction of snake in the box codes,'' \emph{Utilitas Math.}, vol.~40, pp. 97--116, 1991.

\bibitem{Bra14}
M.~Braun, ``{G}ray code for row-reduced echelon forms over the binary field,'' \emph{IEEE Trans.~Inform.~Theory}, vol.~61, no.~2, pp. 829--835, 2014.

\bibitem{BulRus96}
B.~Bultena and F.~Ruskey, ``Transition restricted {G}ray codes,'' \emph{Elec.~J.~of Comb.}, vol.~3, no.~1, p. R11, 1996.

\bibitem{DimDvoTomGreSkr09}
D.~Dimitrov, T.~Dvo{\v{r}}{\'a}k, P.~Gregor, and R.~{\v{S}}krekovski, ``{G}ray code compression,'' in \emph{International Workshop on Combinatorial Algorithms}, 2009, pp. 183--193.

\bibitem{EtzPat96b}
T.~Etzion and K.~G. Paterson, ``Near optimal single-track {G}ray codes,'' \emph{IEEE Trans.~Inform.~Theory}, vol.~42, no.~3, pp. 779--789, May 1996.

\bibitem{Gra53}
F.~Gray, ``Pulse code communication,'' March 1953, {U.S.} Patent 2632058.

\bibitem{HilPatBra96}
A.~P. Hiltgen, K.~G. Paterson, and M.~Brandestini, ``Single-track {G}ray codes,'' \emph{IEEE Trans.~Inform.~Theory}, vol.~42, no.~5, pp. 1555--1561, Sep. 1996.

\bibitem{HilPat01}
A.~P. Hiltgen and K.~G. Patterson, ``Single-track circuit codes,'' \emph{IEEE Trans.~Inform.~Theory}, vol.~47, no.~6, pp. 2587--2595, Sep. 2001.

\bibitem{Hol17}
A.~E. Holroyd, ``Perfect snake-in-the-box codes for rank modulation,'' \emph{IEEE Trans.~Inform.~Theory}, vol.~63, no.~1, pp. 104--110, Jan 2017.

\bibitem{HooRecSawWon15}
S.~Hood, D.~Recoskie, J.~Sawada, and D.~Wong, ``Snakes, coils, and single-track circuit codes with spread {$k$},'' \emph{J.~Comb.~Optim.}, vol.~30, no.~1, pp. 42--62, Jul. 2015.

\bibitem{HorEtz14}
M.~Horovitz and T.~Etzion, ``Constructions of snake-in-the-box codes for rank modulation,'' \emph{IEEE Trans.~Inform.~Theory}, vol.~60, no.~11, pp. 7016--7025, Nov. 2014.

\bibitem{Kau58}
W.~H. Kautz, ``Unit-distance error-checking codes,'' \emph{IRE Transactions on Electronic Computers}, no.~2, pp. 179--180, 1958.

\bibitem{Kle67}
V.~Klee, ``A method for constructing circuit codes,'' \emph{J.~of the ACM}, vol.~14, no.~3, pp. 520--528, 1967.

\bibitem{Mut22}
T.~M{\"u}tze, ``Combinatorial {G}ray codes -- an updated survey,'' \emph{arXiv preprint arXiv:2202.01280}, 2022.

\bibitem{PatTul98}
K.~G. Paterson and J.~Tuliani, ``Some new circuit codes,'' \emph{IEEE Trans.~Inform.~Theory}, vol.~44, no.~3, pp. 1305--1309, May 1998.

\bibitem{Sav97}
C.~D. Savage, ``A survey of combinatorial {G}ray codes,'' \emph{SIAM Rev.}, vol.~39, no.~4, pp. 605--629, Dec. 1997.

\bibitem{Sch14a}
M.~Schwartz, ``Gray codes and enumerative coding for vector spaces,'' \emph{IEEE Trans.~Inform.~Theory}, vol.~60, no.~1, pp. 271--281, Jan. 2014.

\bibitem{SchEtz99}
M.~Schwartz and T.~Etzion, ``The structure of single-track {G}ray codes,'' \emph{IEEE Trans.~Inform.~Theory}, vol.~45, no.~7, pp. 2383--2396, Nov. 1999.

\bibitem{Sin66}
R.~C. Singleton, ``Generalized snake-in-the-box codes,'' \emph{IEEE Trans.~Ele.~Comput.}, vol. EC-15, no.~4, pp. 596--602, Aug. 1966.

\bibitem{Sla79}
P.~J. Slater, ``Open problem,'' in \emph{Proceedings of the 10th Southeastern Conference on Combinatorics, Graph Theory, and Computing}, vol.~2, 1979, pp. 918--919.

\bibitem{Sla89}
------, ``Research problems: problem 109,'' \emph{Discrete Math.}, vol.~76, no.~3, pp. 293--294, 1989.

\bibitem{Sup17}
I.~N. Suparta, ``Some classes of bipartite graphs induced by {G}ray codes,'' \emph{Electron. J. Graph Theory Appl.}, vol.~5, no.~2, pp. 312--324, 2017.

\bibitem{SupVan08}
I.~N. Suparta and A.~J. {van Zanten}, ``A construction of {G}ray codes inducing complete graphs,'' \emph{Discrete Math.}, vol. 308, no.~18, pp. 4124--4132, 2008.

\bibitem{WanFu20}
X.~Wang and F.-W. Fu, ``Snake-in-the-box codes under the {$\ell_\infty$}-metric for rank modulation,'' \emph{Designs, Codes and Cryptography}, vol.~88, pp. 487--503, 2020.

\bibitem{WilErn02}
E.~L. Wilmer and M.~D. Ernst, ``Graphs induced by {G}ray codes,'' \emph{Discrete Math.}, vol. 257, no. 2-3, pp. 585--598, 2002.

\bibitem{WilBla06}
B.~Wilson and M.~Blaum, ``Skew-tolerant {G}ray codes,'' in \emph{Proceedings of the 2006 IEEE International Symposium on Information Theory (ISIT2006), Seattle, WA, USA}, Jul. 2006, pp. 1890--1894.

\bibitem{YehSch12b}
Y.~Yehezkeally and M.~Schwartz, ``Snake-in-the-box codes for rank modulation,'' \emph{IEEE Trans.~Inform.~Theory}, vol.~58, no.~8, pp. 5471--5483, Aug. 2012.

\bibitem{YehSch17}
------, ``Limited-magnitude error-correcting {G}ray codes for rank modulation,'' \emph{IEEE Trans.~Inform.~Theory}, vol.~63, no.~9, pp. 5774--5792, Sep. 2017.

\bibitem{ZhaGe16a}
Y.~Zhang and G.~Ge, ``Snake-in-the-box codes for rank modulation under {K}endall's {$\tau$}-metric,'' \emph{IEEE Trans.~Inform.~Theory}, vol.~62, no.~1, pp. 151--158, Jan. 2016.

\bibitem{ZhaGe16b}
------, ``Snake-in-the-box codes for rank modulation under {K}endall's {$\tau$}-metric in {$S_{2n+2}$},'' \emph{IEEE Trans.~Inform.~Theory}, vol.~62, no.~9, pp. 4814--4818, Sep. 2016.

\end{thebibliography}

\appendix
\subsection{Base case for Construction~\ref{const:skip1general}, $N_0=6$, $L=4$, $R=1$}

\[
\ceil*{\begin{array}{c}
0 0 0 0 0 0  \\
0 0 0 0 0 1  \\
0 0 0 0 1 1  \\
0 0 0 1 1 1  \\ 
0 0 1 1 1 1  \\ 
0 1 1 1 1 1  \\ 
1 1 1 1 1 1  \\ 
1 0 1 1 1 1  \\ 
1 0 0 1 1 1  \\ 
1 0 0 0 1 1  \\ 
1 0 0 0 0 1  \\ 
1 0 0 0 0 0 
\end{array}}
\quad
\abs*{\begin{array}{c}
1 0 0 0 1 0  \\
1 0 0 1 1 0  \\ 
1 0 1 1 1 0  \\ 
1 1 1 1 1 0  \\ 
0 1 1 1 1 0  \\ 
0 0 1 1 1 0  \\ 
0 0 0 1 1 0  \\ 
0 0 0 0 1 0  \\ 
0 0 1 0 1 0  \\ 
0 1 1 0 1 0  \\ 
0 1 0 0 1 0  \\ 
0 1 0 1 1 0
\end{array}}
\quad
\abs*{\begin{array}{c}
0 1 0 1 0 0  \\
0 1 0 0 0 0  \\
0 1 1 0 0 0  \\ 
0 0 1 0 0 0  \\ 
1 0 1 0 0 0  \\ 
1 1 1 0 0 0  \\ 
1 1 0 0 0 0  \\ 
1 1 0 1 0 0  \\ 
1 1 0 1 1 0  \\ 
1 1 0 1 1 1  \\ 
1 1 0 1 0 1  \\ 
1 1 0 0 0 1
\end{array}}
\quad
\floor*{\begin{array}{c}
1 1 1 0 0 1  \\
1 0 1 0 0 1  \\
0 0 1 0 0 1  \\ 
0 1 1 0 0 1  \\ 
0 1 0 0 0 1  \\ 
0 1 0 1 0 1  \\ 
0 1 0 1 1 1  \\ 
0 1 0 0 1 1  \\ 
0 1 1 0 1 1  \\ 
0 0 1 0 1 1  \\ 
1 0 1 0 1 1  \\ 
\end{array}}
\]

\subsection{Base case for Construction~\ref{const:skip1general}, $N_0=7$, $L=3$, $R=3$}

\[
\ceil*{\begin{array}{c}
0 0 0 0 0 0 0  \\
0 0 0 0 0 0 1  \\
0 0 0 0 0 1 1  \\ 
0 0 0 0 1 1 1  \\ 
0 0 0 1 1 1 1  \\ 
0 0 0 1 0 1 1  \\ 
0 0 0 1 0 0 1  \\ 
0 0 0 1 0 0 0  \\ 
0 0 0 1 0 1 0  \\ 
0 0 0 1 1 1 0  \\ 
0 0 0 0 1 1 0  \\ 
0 0 1 0 1 1 0  \\ 
0 1 1 0 1 1 0  \\ 
1 1 1 0 1 1 0  \\ 
1 0 1 0 1 1 0  \\ 
1 0 0 0 1 1 0  \\ 
1 0 0 1 1 1 0  \\ 
1 0 1 1 1 1 0  \\ 
1 1 1 1 1 1 0  \\ 
1 1 0 1 1 1 0  \\ 
1 1 0 0 1 1 0  \\ 
1 1 0 0 0 1 0  \\ 
1 1 0 1 0 1 0  \\ 
1 1 1 1 0 1 0  \\ 
1 0 1 1 0 1 0  \\
0 0 1 1 0 1 0  \\ 
0 1 1 1 0 1 0
\end{array}}
\quad
\abs*{\begin{array}{c}
0 1 0 1 0 1 0  \\ 
0 1 0 0 0 1 0  \\ 
0 1 0 0 1 1 0  \\ 
0 1 0 0 1 0 0  \\ 
0 1 0 0 0 0 0  \\ 
0 1 0 1 0 0 0  \\ 
0 1 1 1 0 0 0  \\ 
0 1 1 0 0 0 0  \\ 
0 1 1 0 1 0 0  \\ 
0 1 1 1 1 0 0  \\ 
0 1 0 1 1 0 0  \\ 
0 0 0 1 1 0 0  \\ 
0 0 1 1 1 0 0  \\ 
0 0 1 0 1 0 0  \\ 
0 0 1 0 0 0 0  \\ 
0 0 1 0 0 1 0  \\ 
0 0 1 0 0 1 1  \\ 
0 0 1 0 0 0 1  \\ 
0 0 1 0 1 0 1  \\ 
0 0 1 1 1 0 1  \\ 
0 0 0 1 1 0 1  \\ 
0 1 0 1 1 0 1  \\ 
0 1 1 1 1 0 1  \\
0 1 1 0 1 0 1  \\ 
0 1 1 0 0 0 1  \\ 
0 1 1 1 0 0 1  \\ 
0 1 0 1 0 0 1
\end{array}}
\quad
\abs*{\begin{array}{c}
0 1 0 0 0 0 1  \\ 
0 1 0 0 1 0 1  \\ 
0 1 0 0 1 1 1  \\ 
0 1 0 0 0 1 1  \\ 
0 1 0 1 0 1 1  \\ 
0 1 1 1 0 1 1  \\ 
0 0 1 1 0 1 1  \\ 
1 0 1 1 0 1 1  \\ 
1 1 1 1 0 1 1  \\ 
1 1 0 1 0 1 1  \\ 
1 1 0 0 0 1 1  \\ 
1 1 0 0 1 1 1  \\ 
1 1 0 1 1 1 1  \\ 
1 1 1 1 1 1 1  \\ 
1 0 1 1 1 1 1  \\ 
1 0 0 1 1 1 1  \\ 
1 0 0 0 1 1 1  \\ 
1 0 0 0 0 1 1  \\ 
1 0 0 0 0 0 1  \\ 
1 0 0 0 1 0 1  \\ 
1 0 0 1 1 0 1  \\
1 0 1 1 1 0 1  \\ 
1 0 1 0 1 0 1  \\ 
1 0 1 0 0 0 1  \\ 
1 0 1 0 0 1 1  \\ 
1 0 1 0 0 1 0  \\ 
1 0 1 0 0 0 0 
\end{array}}
\quad
\floor*{\begin{array}{c}
1 0 1 0 1 0 0  \\ 
1 0 1 1 1 0 0  \\ 
1 0 0 1 1 0 0  \\ 
1 0 0 0 1 0 0  \\ 
1 0 0 0 0 0 0  \\ 
1 0 0 1 0 0 0  \\ 
1 0 1 1 0 0 0  \\ 
1 1 1 1 0 0 0  \\ 
1 1 0 1 0 0 0  \\ 
1 1 0 0 0 0 0  \\ 
1 1 0 0 1 0 0  \\ 
1 1 0 1 1 0 0  \\ 
1 1 1 1 1 0 0  \\ 
1 1 1 0 1 0 0  \\ 
1 1 1 0 0 0 0  \\ 
1 1 1 0 0 1 0  \\ 
1 1 1 0 0 1 1  \\ 
1 1 1 0 0 0 1  \\ 
1 1 1 0 1 0 1  \\
1 1 1 1 1 0 1  \\ 
1 1 0 1 1 0 1  \\ 
1 1 0 0 1 0 1  \\ 
1 1 0 0 0 0 1  \\ 
1 1 0 1 0 0 1  \\ 
1 1 1 1 0 0 1  \\ 
1 0 1 1 0 0 1  \\ 
0 0 1 1 0 0 1
\end{array}}
\]

\end{document}